\documentclass[twocolumn,nofootinbib]{revtex4-1}
\usepackage{latexsym,epsfig,amssymb, amsmath,nicefrac}
\usepackage{epsfig,graphicx}
\usepackage[usenames, dvipsnames]{color}

\newcommand{\vt}{\vartheta}

              \newcommand{\rf}[1]{(\ref{#1})}

\def\bfone{\relax{\rm 1\kern-.35em 1}}

\newcommand{\be}{\begin{equation}}
\newcommand{\ee}{\end{equation}}
\newcommand{\ben}{\begin{displaymath}}
\newcommand{\een}{\end{displaymath}}
\newcommand{\bea}{\begin{eqnarray}}
\newcommand{\eea}{\end{eqnarray}}

\newcommand{\bean}{\begin{eqnarray*}}
\newcommand{\eean}{\end{eqnarray*}}

\newcommand{\vp}{\varphi}

\def\K{K{\"a}hler}

\makeatletter \@addtoreset{equation}{section} \makeatother

\begin{document}

\title{\Large{Cosmological  Attractors and Initial Conditions for Inflation}}

\author{John Joseph M. Carrasco, Renata Kallosh and Andrei Linde}

\affiliation{Department of Physics and SITP, Stanford University, \\ 
Stanford, California 94305 USA}

\begin{abstract}
Inflationary  $\alpha$-attractor models in supergravity, which provide excellent fits to the latest observational data, are based on the Poincar\'e disk hyperbolic geometry. 
 We refine these models by constructing  \K\,  potentials with built-in inflaton shift symmetry and by making a canonical choice of the goldstino \K\, potential.  The refined models are stable with respect to all scalar fields at all  $\alpha$, no additional stabilization terms are required. The scalar potential $V$ has a nearly Minkowski minimum at small values of the inflaton field $\vp$, and an infinitely long dS valley of constant depth and width at large $\vp$. Because of the infinite length of this shift-symmetric valley, the initial value of the inflaton field at the Planck density  is expected to be extremely large. We show that the inflaton field $\vp$ does not change much until all fields lose their energy and fall to the bottom of the dS valley at large $\vp$. 
This provides natural initial conditions for inflation driven by the inflaton field slowly rolling along the dS valley towards the minimum of the potential at small $\vp$. 
A detailed description of this process is given for $\alpha$-attractors in supergravity, but we believe that our general conclusions concerning naturalness of initial conditions for inflation are valid for a broad class of inflationary models with sufficiently flat potentials.

\end{abstract}

\maketitle

\smallskip


\section{Introduction}\label{intro}

During the last two years, a new class of inflationary theories have been discovered: ``cosmological attractors.'' This class is very broad, including conformal attractors \cite{Kallosh:2013hoa}, universal attractors with non-minimal coupling to gravity \cite{Kallosh:2013tua}, and $\alpha$-attractors \cite{Ferrara:2013rsa,Kallosh:2013yoa,Cecotti:2014ipa,Kallosh:2015lwa}, and it also incorporates many previously existing models such as Starobinsky model \cite{Starobinsky:1980te}, GL model \cite{Goncharov:1983mw,Linde:2014hfa}, and Higgs inflation model \cite{Salopek:1988qh}.  Despite very different origins, all of these models make very similar cosmological predictions providing an excellent match to the latest cosmological data  \cite{Ade:2015tva,Planck:2015xua}. Moreover, these models can be further extended to describe not only inflation, but also the theory of dark energy/cosmological constant and supersymmetry breaking \cite{Ferrara:2014kva}.

On a purely phenomenological level, the main features of inflation in all of these models can be represented in terms of a single-field toy model with the Lagrangian  \cite{Galante:2014ifa,Kallosh:2015zsa}
 \be
 {1\over \sqrt{-g}} \mathcal{L}_{\rm T} = {1\over 2}   R - {1\over 2} {\partial \phi^2\over (1-{\phi^{2}\over 6\alpha})^{2}}  - V(\phi)   \,  .
\label{cosmo}\ee
Here $\phi(x)$ is the scalar field, the inflaton. The parameter  $\alpha$  can take any positive value. In the limit $\alpha \rightarrow \infty$ this model coincides with the standard chaotic inflation models with a canonically normalized field $\phi$ and the inflaton potential $V(\phi)$  \cite{Linde:1983gd}. 

However, the field $\phi$ in \rf{cosmo} is not canonically normalized. It must satisfy the condition $\phi^2<6\alpha$, for the sign of the inflaton kinetic term to remain positive.  One can easily go to canonically normalized variables $\vp$ by solving the equation ${\partial \phi\over 1-{\phi^{2}\over 6\alpha}} = \partial\vp$, which yields
\be 
\phi = \sqrt {6 \alpha}\, \tanh{\varphi\over\sqrt {6 \alpha}} \ .
\ee
In terms of these variables, a tiny vicinity of the singular boundary of the moduli space at $\phi=\sqrt{6\alpha}$ stretches and extends to infinitely large $\vp$. As a result, generic potentials $V(\phi) = V(\sqrt {6 \alpha}\, \tanh{\varphi\over\sqrt {6 \alpha}})$ at large $\vp$ approach an infinitely long dS inflationary plateau with the height corresponding to the value of $V(\phi)$ at the boundary:
\be
V_{dS} = V(\phi)|_{\phi = \pm \sqrt {6 \alpha}} \ .
\ee  

This universal origin of the shape of these potentials leads to universality in their predictions, as explained in  \cite{Kallosh:2013hoa,Ferrara:2013rsa,Kallosh:2013yoa} and formulated in a particularly general way in \cite{Galante:2014ifa}:  The kinetic term in this class of models has a pole at the boundary of the moduli space. If inflation occurs in a vicinity of such a pole, and the potential near the pole can be well represented by its value and its first derivative near the pole, all other details of the potential far away from the pole (from the boundary of the moduli space) become unimportant for making cosmological predictions. In particular, the spectral index depends solely on the order of the pole, and the tensor-to-scalar ratio relies on the residue  \cite{Galante:2014ifa}. All the rest is practically irrelevant, as long as the field after inflation falls into a stable minimum of the potential, with a tiny value of the vacuum energy, and stays there.

This new class of models accomplishes for inflationary theory something similar to what inflation does for cosmology. Inflation stretches the universe making it flat and homogeneous, and the structure of the observable part of the universe becomes very stable with respect to the choice of initial conditions in the early universe. Similarly, stretching of the moduli space near its boundary upon transition to canonical variables makes inflationary potentials very flat and results in predictions which are very stable with respect to the choice of the inflaton potential.

The present work will pursue two different goals simultaneously. First of all, we will study geometrical properties of the moduli space in the supergravity realizations of these models, following  \cite{Kallosh:2015zsa,Carrasco:2015uma}. We will reformulate these models in terms of \K\ potentials and field variables which keep their geometric properties manifest. This new formulation will allow us to approach the problem of initial conditions for inflation in these models in a novel, more transparent way. 

The problem of initial conditions in these models is not quite trivial. In the simplest chaotic inflation models such as ${m^{2}\over 2}\phi^{2}$ inflation may start very close to the Planck density. According to \cite{Linde:1983gd,Linde:1985ub,Linde:1986fd,Linde:1984ir,Vilenkin:1984wp,Linde:2005ht}, this makes initial conditions for inflation quite natural. However, in the new class of models discussed above, as well as in the Starobinsky model and Higgs inflation, the inflationary regime begins at the energy density 10 orders below the Planck energy density.  A solution of the problem of initial conditions in such models was discussed in \cite{Linde:2014nna}. Here we will revisit it; we will show how this problem can be solved in the supergravity realizations of $\alpha$-attractors. Most of our conclusions will have more general validity, being  applicable, in particular, to generic non-supersymmetric attractor models \rf{cosmo}. We will also show that in some cases, such as supergravity $\alpha$-attractors with $\alpha\ll 1$, inflation can begin at the density approaching  the Planck density, thus reducing the problem of initial conditions to the one already addressed in \cite{Linde:1983gd,Linde:1985ub,Linde:1986fd,Linde:1984ir,Vilenkin:1984wp,Linde:2005ht}.

 There are two types of technical improvements of our $\alpha$-models which we will develop in this paper. The first one, following \cite{Kallosh:2015zsa,Carrasco:2015uma}, will  allow us to use the \K\, frame where the inflaton shift symmetry is present in the new \K\, potentials. The second improvement with respect to earlier models corresponds to changing the  \K\, potential for the goldstino multiplet, making it canonical rather than part of the logarithmic structure, which has a consequence: an improved {\em manifest} stability.

We will make a choice of variables in which the inflaton forms a  Killing direction of the moduli space geometry.  Namely, our holomorphic disk variable $Z$ and the half-plane variable $T$   used in  \cite{Kallosh:2013hoa,Kallosh:2013yoa,Cecotti:2014ipa} will be represented  by the Killing adapted moduli space coordinates
\be
Z= {T-1\over T+1}= \tanh
 {\vp+i \vt \over \sqrt {6\alpha}}  \ .
\label{vpt}\ee
Here the inflaton $\vp$ and the orthogonal field $\vt$ form a geometry independent on a Killing direction $\vp$:
\be g_{\vp\vp} (\vt)= g_{\vt\vt}(\vt) = {1 \over  \cos^{2} \sqrt{2  \over 3\alpha} \vartheta}  \ee
As a result, the time evolution in our models with initial large kinetic energy, when the role of the potential is negligible, will be controlled by the fact that the momentum in the inflaton direction is preserved, namely
\be
\dot P_{\vp} =0 \qquad {\rm where} \qquad P_{\vp}= a^3(t) g_{\vp\vp} (\vt) \dot \vp
\ee
This  geometric fact helps us to argue that 
the total shift of the field $\vp$ due to its initial velocity is about 10 Planck units or less, after which all memory about the initial velocity of the field $\vp$ at the Planckian time completely disappears.

We will also numerically solve the Friedmann equations in FRW space-time metric for generic initial values of $\vt, \vp, \dot \vt, \dot \vp$  confirming our analytic analysis: we have an inflationary attractor behavior, where the memory about initial values of $\vt, \vp, \dot \vt, \dot \vp$  disappears and period of slow-roll inflation at the minimum of the potential at $\vt=0$ takes place.

We will show that with new \K\, potentials which have the inflaton shift symmetry in $Z$ or $T$ variables, the superpotentials are  simpler and the relation between models in disk and half-plane variables  simplifies. 

We will than proceed with the analysis of initial conditions for inflation in these models, with our new choice of variables, by making choices of initial values of the inflaton and its partner and by studying the time evolution of these fields, before and during inflation. The geometric nature of our models, and the existence of infinite dS valleys of constant width in our potentials, help to resolve this issue and allowing us to argue that the vast majority of initial conditions in these models leads to successful inflation.

\section{From disk to half-plane: New \K\ potentials}
The cosmological attractor models can be described either in disk or in half-plane variables \cite{Cecotti:2014ipa,Kallosh:2015zsa}. The corresponding boundary of the moduli space,  which plays an important role in these models, is either at $Z\bar Z<1$, or a half-plane with  $T+\bar T > 0$.

Here we summarize the relation between disk and half-plane variables for generic case of 2-superfield models with our choice of the \K\, potentials and most general superpotentials. 

The relation between the \K\ potentials and superpotentials in the disk and half-plane variables requires a simple Caley transform, as suggested in \cite{Cecotti:2014ipa} 
\be
Z= {T-1\over T+1}\, ,    \qquad T= {1+Z\over 1-Z}\, .
\label{Z}
\ee 
We will represent the \K\ potential in the following form: 
\be\label{NewD}
K_{\mathbb{D}}= -{3  \alpha\over 2}  \log  \left[ {( 1- Z \bar{Z} )^2  \over  (1-Z^2) (1- \bar{Z}^2)}  \right] + S \bar{S} \, ,  
\ee
\be
W_{\mathbb{D}}= A(Z)  + S\, B(Z)\,   \ .
\ee
where $S$ is a  supermultiplet with a goldstino fermion and a sgoldstino scalar. This field may either belong to the   usual unconstrained chiral multiplet, or it may be a nilpotent superfield as studied in \cite{Ferrara:2014kva}. We will discuss both options in this paper. 

We present our models as functions of a  complex variable
\be
\Phi \equiv \vp +i \vt \ ,
\ee
where $\vp$ will be the inflaton in cosmological models and $\vt$ will describe the orthogonal direction and 
\be
Z= \tanh
 {\Phi \over \sqrt {6\alpha}}  \ .
\label{Phi}\ee
Our \K\, potential \rf{NewD} in these variables has a manifest inflaton shift symmetry, $\vp '= \vp+c$
\be
K_{\mathbb{D}}(Z, \bar Z)  \quad \Rightarrow \quad K= -3\alpha \log \Big [\cosh \Big ({\Phi-\bar \Phi \over \sqrt{6\alpha}}\Big) \Big] + S \bar{S} \ .
\label{JJ1}\ee
The superpotential is now
\be
W_{\mathbb{D}} \quad \Rightarrow \quad W= A \Big (\tanh
 {\Phi \over \sqrt {6\alpha}}
 \Big)  + S\, B \Big (\tanh
 {\Phi \over \sqrt {6\alpha}}
\Big)
\,   \ .
\ee
Note that in our models $\vt=0$ during inflation and therefore the new holomorphic variable $\Phi$ during inflation becomes a real canonical variable. This is also easy to see from  the kinetic terms in these variables, which are conformal to flat,
\be
 ds^2=    {d\vp^2 + d\vartheta^2 \over 2\cos^{2}\sqrt{2  \over 3\alpha} \vartheta }= {1\over 2} g_{\vp\vp} d\vp^2 +  {1\over 2} g_{\vt\vt} d\vt^2 \ .
\label{JJ}\ee
At $\vt=0$ they are canonical
\be
 ds^2|_{\vt=0} =    {d\vp^2 + d\vartheta^2 \over 2 } \ .
\label{JJa}\ee
We can also use the  half-plane variables $T+\bar{T}>0$  where we have
 \be\label{half}
K_{\mathbb{H}}= -{3  \alpha\over 2}  \log\left[ { \left(T + \bar{ T}    \right) ^2\over  4 \, T\, \bar{T}  } \right]  + S \bar{S} ,  
\ee
\be W_{\mathbb{H}}= G(T)+ S F(T)\ .
\ee
Now the disk and the half-plane models are  related simply by the Caley transform \rf{Z}, so that transition from one picture to the other is a simple substitution
\be
K_{\mathbb{D}} \Big (Z= {T-1\over T+1}, \, \bar{Z}= {\bar{T}-1\over \bar{T}+1} \Big ) =  K_{\mathbb{H}} (T, \bar{ T})  \ .
\ee
and 
\be
W_{\mathbb{D}} \Big (Z= {T-1\over T+1} , S\Big ) =  W_{\mathbb{H}} (T )  \ .
\ee
This also means that 
\be
  G (T )= A\Big (Z= {T-1\over T+1} \Big ) \, , 
\quad 
 F (T ) = B\Big (Z= {T-1\over T+1} \Big )  \ .
\ee
When we hold $S\bar S$ outside of the $\log$ part of the \K\, potential, the field $S$ does not change from one picture to the other. However, for any models with $S\bar S$ inside the $\log$ part of the \K\, the potential which we used before, 
 the relation between  the goldstino multiplets in $Z$ and $T$ variables  involves the dependence on the inflaton superfield,  as shown in \cite{Cecotti:2014ipa}. We will explain below that  when the field $S$ is outside the $\log$ in the \K\, potential the inflaton partner is stable for any $\alpha$. Therefore we will focus here on models with canonical \K\, potentials for the $S$ field as shown in eqs. \rf{NewD}, \rf{JJ1} and \rf{half}.

\section{$\alpha$-attractors and their stability} 


We will begin with a rather simple and general class of $\alpha$-attractors in disk variables, with the \K\ potential \rf{NewD}
 and superpotential 
\be
W_{\mathbb{D}}   = \sqrt{\alpha} \,  \mu \ \, S   \,  f(Z)\, .
\label{gen}\ee
Investigation of this theories simplifies considerably if during and after inflation the field $S$ vanishes, along with the imaginary part of the field $Z$. Indeed, as explained in \cite{Carrasco:2015uma}, the \K\ potential \rf{NewD} has a shift symmetry under the shift of the inflaton field during inflation, when $x= {\rm Im}\, Z = 0$: The \K\ potential vanishes independently of the value of the inflaton field $z = {\rm Re}\, Z$.

In that case, one can show that the potential of the canonically normalized inflaton field $\vp$, which is defined by the relation $z = \tanh{ \vp\over \sqrt {6\alpha}}$, is given by 
\be
V = \alpha \mu^2 f^{2}(z)= \alpha \mu^2 f^2 \bigl(\tanh {\vp\over \sqrt {6\alpha}}\bigr)  .
\ee
The potential has an infinitely long dS plateau at $\vp \gg \alpha$, exponentially rapidly approaching its asymptotic value
\be
V_{dS} = \alpha \mu^2 \ .
\ee
Predictions from such theories provide a very good fit to observational data for a broad class of functions $f(Z)$ as discussed in \cite{Kallosh:2013hoa,Kallosh:2013yoa}.

However, for such analysis to hold, it is important to verify that $S = s \, e^{i\,\gamma}=0$, and $x= {\rm Im}\, Z = 0$, or to find a way to stabilize these fields at their zero values. 
The point $S = x =0$ is indeed an extremum of the potential for $S$ and $x$, but one should also check whether this extremum is a minimum, or a maximum of the potential.

Let us start with the field $x$. One can show that its mass squared is given by 
\be\label{mx}
m^{2}_{x}(z) = {V_{dS}\over 3} (6\alpha f^{2}(z) +(1-z^{2})^{2} [(f'(z))^{2} - f(z) f''(z)]) \ .
\ee
If we consider potentials  $V=  \alpha \mu^2 f^{2}(z)$ vanishing at $z = 0$, then at  the minimum one has $f(0) = 0$, and $m^{2}_{x}(0)$ is positive and coincides with the inflaton mass squared at that point, 
\be
m^{2}_{x}(0) =m^{2}_{z}(0) ={1\over 3} [(f'(0))^{2}]) \ .
\ee
Meanwhile \rf{mx} implies that during inflation  
\be\label{mass}
m^{2}_{x}(z) = 2V(z) = 6H^{2}, 
\ee
up to small corrections suppressed by the slow-roll parameters. Thus during inflation this field is strongly stabilized for all values of $\alpha$ and tends to rapidly roll down to $x = 0$ and stay there.

The condition $S=0$ can be satisfied e.g. if the goldstino superfield is nilpotent \cite{Ferrara:2014kva}, or if it is strongly stabilized at $S = 0$ by terms $\sim (S\bar S)^{2}$ which can be added to the \K\ potential \cite{Kallosh:2010ug}, \cite{Kallosh:2013yoa}. Indeed, one of these two options is compulsory in many theories where the \K\ potential has the term $S\bar S$ under the logarithm. However, in the theories with the \K\ potential \rf{NewD} considered in this section the situation is much better: The mass squared of the field $S = s \, e^{i\,\gamma}$ at $s = x = 0$ does not depend on $\gamma$ and is given by  
\be\label{ms}
m^{2}_{s}(z) = {V_{dS}\over 9} (1-z^{2})^{4} (f'(z))^{2} \geq 0 \ .
\ee
This mass becomes much smaller than $H$ during inflation at $z \approx 1$, but the point $s = 0$ is always the minimum of the potential with respect to the field $S$. This field cannot change much because of the overall fast growing factor $e^{s^{2}}$ in its potential. Thus, the classical field $s$ rolls down towards $s = 0$ and stays there even without adding any stabilization terms to the theory. Since its mass is small, it can experience small inflationary perturbations in the vicinity of $s =0$, but these perturbations are inconsequential unless one makes some radical assumptions about strong suppression of the probability of decay of the field $S$ in the process of reheating \cite{Demozzi:2010aj}.

The same models can be discussed in terms of half-plane variables, with $Z \Rightarrow  {T-1\over T+1}$. Thus, one can use the \K\ potential \rf{half} and
\be
W^{\text{general-T }}_{\mathbb{H}}  \equiv  \sqrt{\alpha}  \, \mu \, S \, \,   f \Big ({ T- 1  \over T+1}\Big )\, .
\ee
In what follows we will give some simple examples of these models, both in disk variables and in half-plane ones.

\section{T-models}\label{tt}

The simplest $\alpha$-attractor T-model has \K\ potential \rf{NewD} and superpotential \rf{gen} with $f(Z) = Z$:
\be\label{simpleT}
W^{\,\text{simple-T}}_{\mathbb{D}} =  \sqrt{\alpha}  \, \mu \, S \,  \, Z\,  \ .
\ee
The inflationary potential for the canonical field $\vp$ is
\be
V= \alpha \mu^2 \tanh^2{ \vp\over \sqrt {6\alpha}} \ .
\label{Vsimple}\ee
We show these potentials in Fig.~1 for few values of $\alpha$.

\begin{figure}[ht!]
\centering
{
\includegraphics[scale=.42]{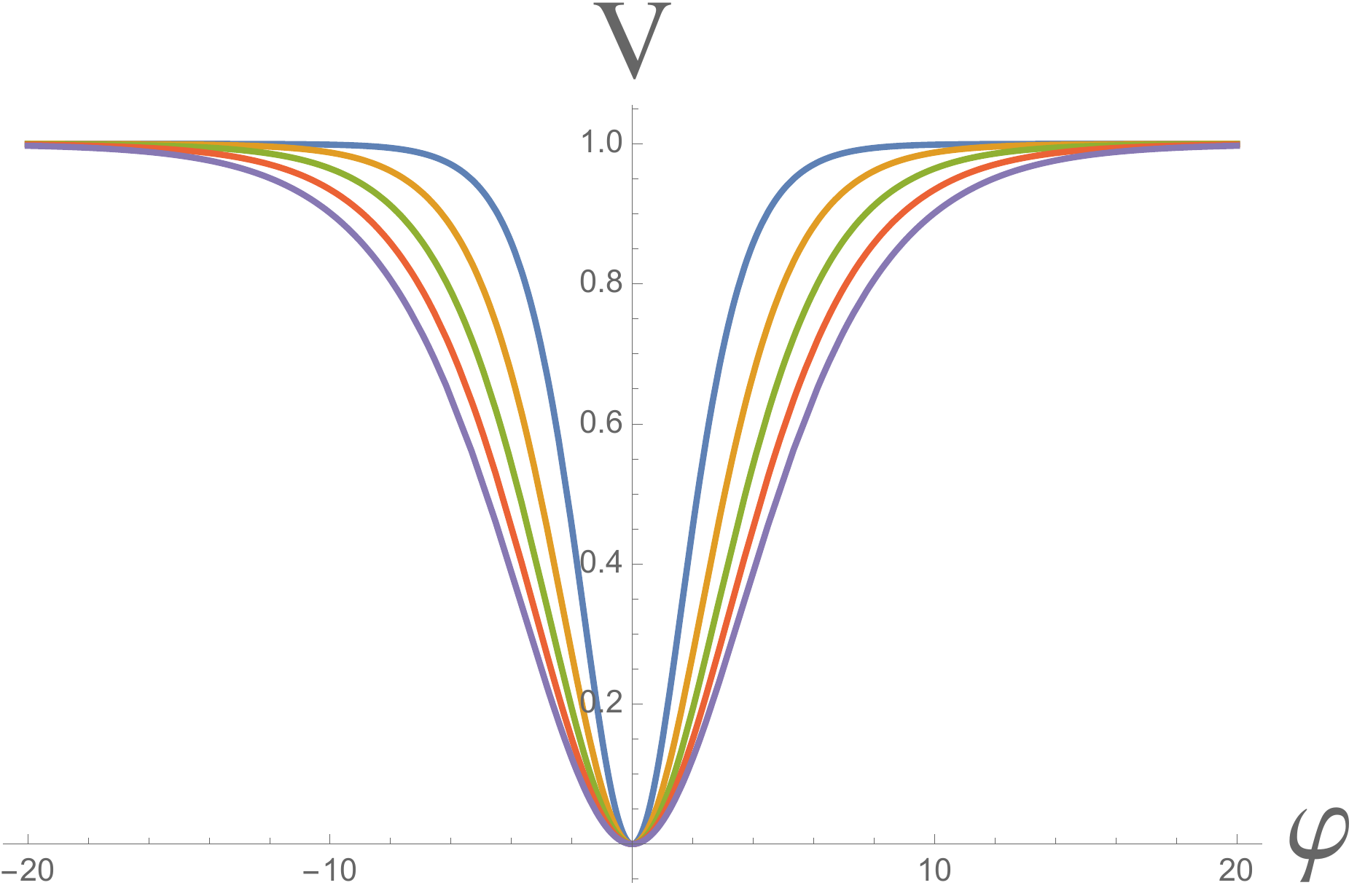}
\label{fig:newlong}}~~~~~
\caption{\footnotesize T-model potential  $V= \alpha \mu^2 \tanh^{2n}{\vp\over \sqrt {6\alpha}}$  for $\alpha = 1,...,5$ at ${\rm Im} \, Z=0$. It is plotted in units $\alpha \mu^2=1$. The potential has two symmetric shoulders.}
\end{figure}
\begin{figure}[ht!]
\centering
{
\includegraphics[scale=.32]{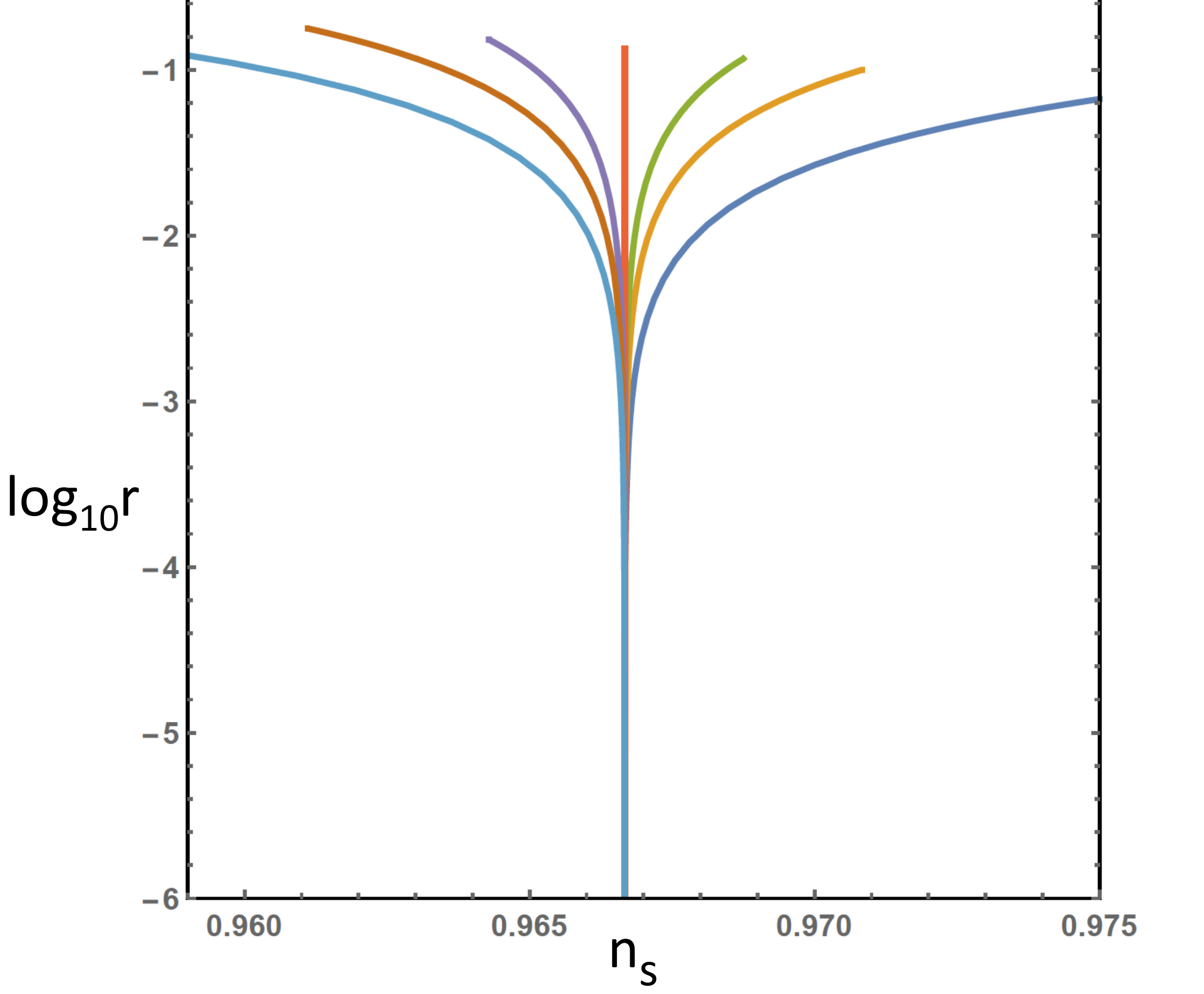}
\label{fig:newlongb}}
\caption{\footnotesize Here we present the cosmological observables $n_s$ and $r$ for simple T-models with different monomial potentials $V=\alpha\mu^{2 }\Big(\tanh (\varphi/\sqrt{6\alpha})\Big )^{2n}$ with $n = (1/2, 3/4, 7/8, 1, 3/2, 2, 3)$  starting from the right, increasing to the left, with the purple line for $n=1$ in the middle. We use the   logarithmic scale in $r$. }
\end{figure}

One may also consider a more general case where $f[Z]= Z^n$ and 
\be
V= \alpha \mu^2 \tanh^{2n}{\vp\over \sqrt {6\alpha}} \ .
\ee
We illustrate the attractor properties of these models in Fig.~2 in the logarithmic scale in $r$. The attractor line, starting at $r=10^{-3}$ towards smaller $r$, remains universal, the value of $r$ depends linearly on $\alpha$  and looses any dependence on $n$: 
\be\label{nsr}
 1 -n_{s} = {2\over N_{e}}\, , \qquad r =  {12\alpha \over N_{e}^{2} } \ ,
 \ee
where $N_{e}$ is the number of e-foldings. In the simplest T-models \rf{Vsimple} the amplitude of the scalar perturbations also does not depend on $\alpha$ and matches the Planck~2015 normalization for $\mu \approx 10^{{-5}}$ \cite{Kallosh:2015lwa}.


Same models in half-plane variables are found by using the \K\ potential \rf{half} and replacing $Z \Rightarrow  {T-1\over T+1}$ in equations of the previous section. In particular, the model with $f(Z) = {Z}$ becomes 
\be
W^{\,\text{simple-T}}_{\mathbb{H}} =  \sqrt{\alpha}  \, \mu \, S \, \,  { T- 1  \over T+1} \,  .
\label{Tsimple}
\ee

\section{E-models}\label{ee}
Another interesting class of models, called E-models, appears if one uses $K_{\mathbb{D}}$  \rf{NewD} and the following superpotentials
\be
W^{\,\text{simple-E}}_{\mathbb{D}} \equiv  \sqrt{\alpha}  \, \mu \, S \,  \, {2 Z\over Z+1}\, .
\ee

The inflationary potential for the canonical field $\vp$ is
\be
V= \alpha \mu^2  \Big (1- e^{-{ \sqrt {2\over 3 \alpha} \vp}}\Big )^{2} .
\ee
We show these potentials in Fig.~3 for few values of $\alpha$.
\begin{figure}[h!]
\centering
\includegraphics[scale=.38]{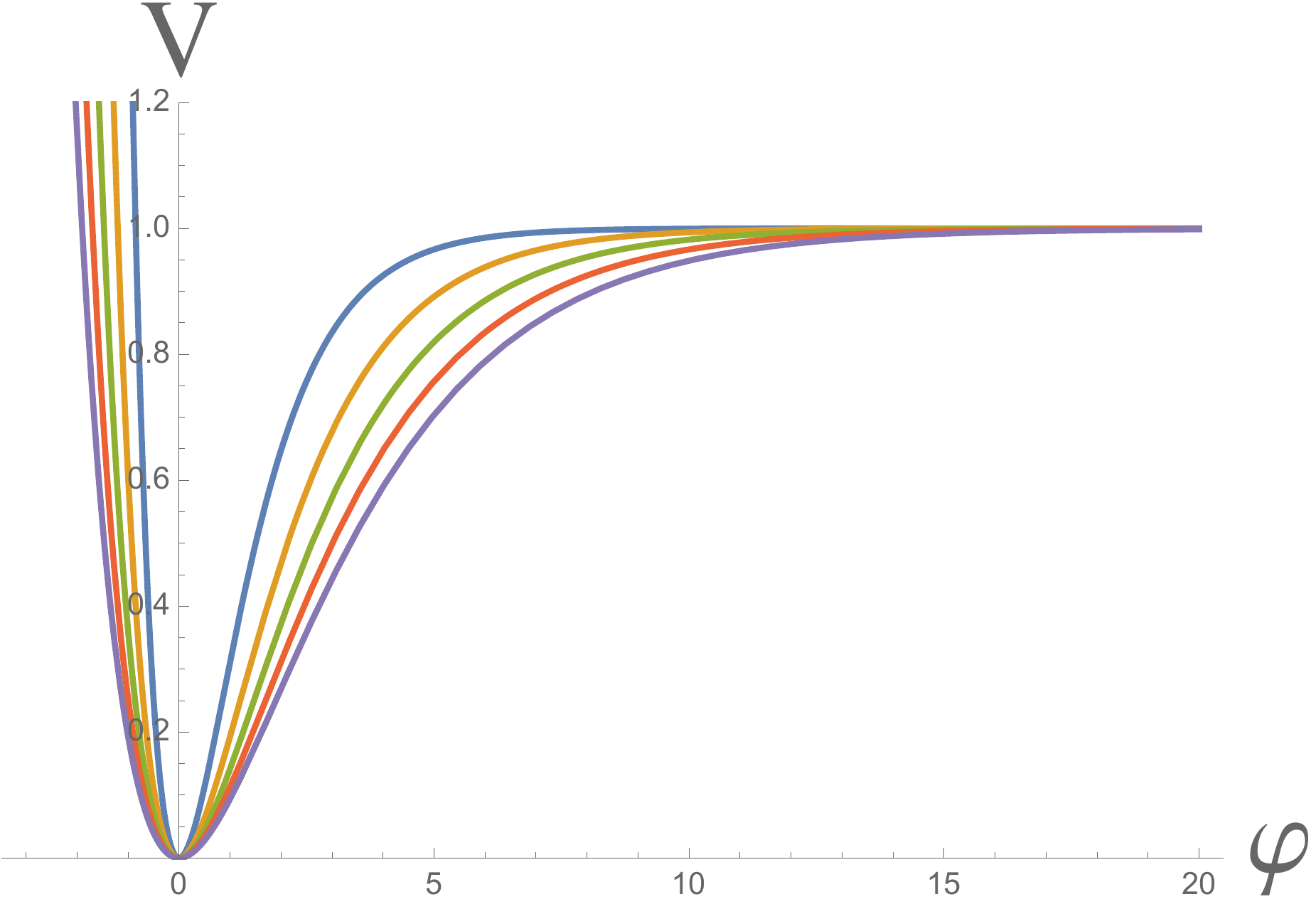}
\caption{\footnotesize Potentials  of simple E-models  $V=\Big (1- e^{-{ \sqrt {2\over 3 \alpha} \vp}}\Big )^{2}$  for $\alpha = 1, 2,3,4,5$. It has an inflationary plateau for large positive $\vp$ and a wall at small negative $\vp$.}
\label{fig:simpleObservables}
\vspace{-.3cm}
\end{figure} 
\begin{figure}[h!]
\centering
\includegraphics[scale=.32]{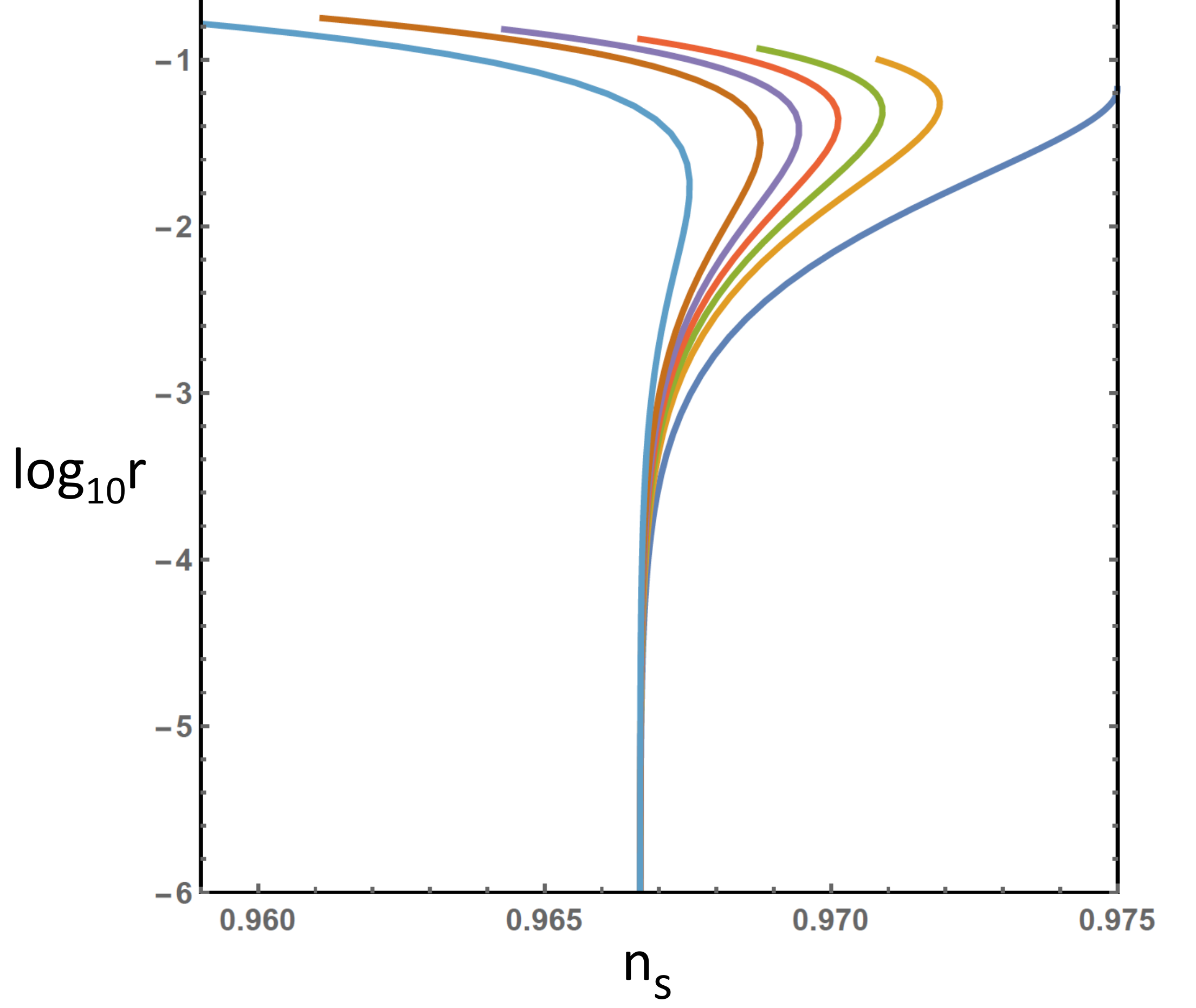}
\caption{\footnotesize  Here we present the cosmological observables $n_s$ and $r$  for  E-models with different monomial potentials $V= (1- e^{-{ \sqrt {2\over 3 \alpha} \vp}})^{2n} $
with $n = (1/2, 3/4, 7/8, 1, 3/2, 2, 3)$  starting from the right, increasing to the left, with the  line for $n=1$ in the middle. The scale in $r$ is  logarithmic.}
\label{fig:simpleObservables2}
\vspace{-.3cm}
\end{figure} 

In more general case
\be
W^{\text{general-E }}_{\mathbb{D}}  \equiv  \, \sqrt{\alpha}  \, \mu \, S \, \,   f \Big ({2 Z\over Z+1} \Big )\, \,
\ee
and the potentials are
\be
V  = \alpha \mu^2 f^2\Big [ (1- e^{-{ \sqrt {2\over 3 \alpha} \vp}}\Big )\Big].
\ee
Let us take a simple example of a general case where $f\big[{2 Z\over Z+1} \big] =\big [{2 Z\over Z+1}\big]^n$ and 
\be
V= \alpha \mu^2 \Big (1- e^{-{ \sqrt {2\over 3 \alpha} \vp}}\Big )^{2n}.
\ee
We illustrate the attractor properties of these models in Fig.~4 using a logarithmic scale in $r$. The attractor line, starting at $r\sim 5\cdot 10^{-4}$ towards smaller $r$, remains universal, the value of $r$ depends on $\alpha$ linearly and becomes completely independent of $n$: $r=\alpha {12\over N^2}$

During inflation, in all of these models, the effective mass of the inflaton partner, taking into account the non-minimal kinetic term, is given by $6H^2$, 
up to slow roll parameters, where $H$ is a Hubble parameter during inflation. Thus in this class of improved $\alpha$-attractor models in supergravity there is no need for stabilization terms.

The models in half-plane variables are found by using the \K\ potential \rf{half} and replacing $Z \Rightarrow  {T-1\over T+1}$ in equations of the previous section. In particular, the model with $f(Z) = {2Z\over Z+1}$ becomes 
\be
W^{\,\text{simple-E}}_{\mathbb{H}} \equiv  \sqrt{\alpha}  \, \mu \, S \,  \, {T-1\over T}\ .
\ee

\section{Asymmetric T-models}\label{aa}

In section \ref{tt}, we considered the simplest T-models which have potentials with two shoulders of equal height. In comparison, E-models have only one inflationary plateau, with the potential infinitely rising at large negative values of the field $\vp$, without forming a second inflationary plateau. In this section we will consider an intermediate class of models, where the potentials have two shoulders with differing heights~\cite{Kallosh:2015lwa,Roest:2015qya,Linde:2015uga}. In a certain sense, one can think about such models as interpolating between the simplest T-models and E-models, but they are different because each of these two asymmetric shoulders may serve as an independent inflationary plateau. Such models offer additional flexibility, which, as we will see, may be quite handy.

Let us use disk \K\ potential  \rf{NewD} and a superpotential in \rf{gen}  with  $f[Z]= (1- c\,Z)$:
\be
W^{\,\text{linear}}_{\mathbb{D}} =  \sqrt{\alpha}  \, \mu \, S \,  \, (1- c\,Z) \ .
\ee
The resulting potential in the inflaton direction in $Z$ variables is  
\be
V= \alpha \mu^2 (1-c\,Z)^{2} \ .
\ee
We will assume for definiteness that $c > 0$. The potential has a minimum at $Z = 1/c$, which belongs to the required range $|Z| < 1$ for $c >1$. In terms of the canonical field $\vp$, the potential is 
\be\label{lin}
V= \alpha \mu^2 \left(c\,\tanh{ \vp\over \sqrt {6\alpha}}-1\right)^{2} \ .
\ee
At large negative values of the inflaton field, the potential has a ``shoulder'' at $V_{-} =(c+1)^{2}$, and at large positive $\vp$ the potential has a shoulder of a different height, $V_{+} = (c-1)^{2}$, so that
\be
{V_{+}\over V_{-}} = \left({c-1\over c+1}\right)^{2} \ .
\ee
This ratio can be arbitrarily small for $0<c-1 \ll  1$.

\begin{figure}[ht!]
\centering
{
\includegraphics[scale=.45]{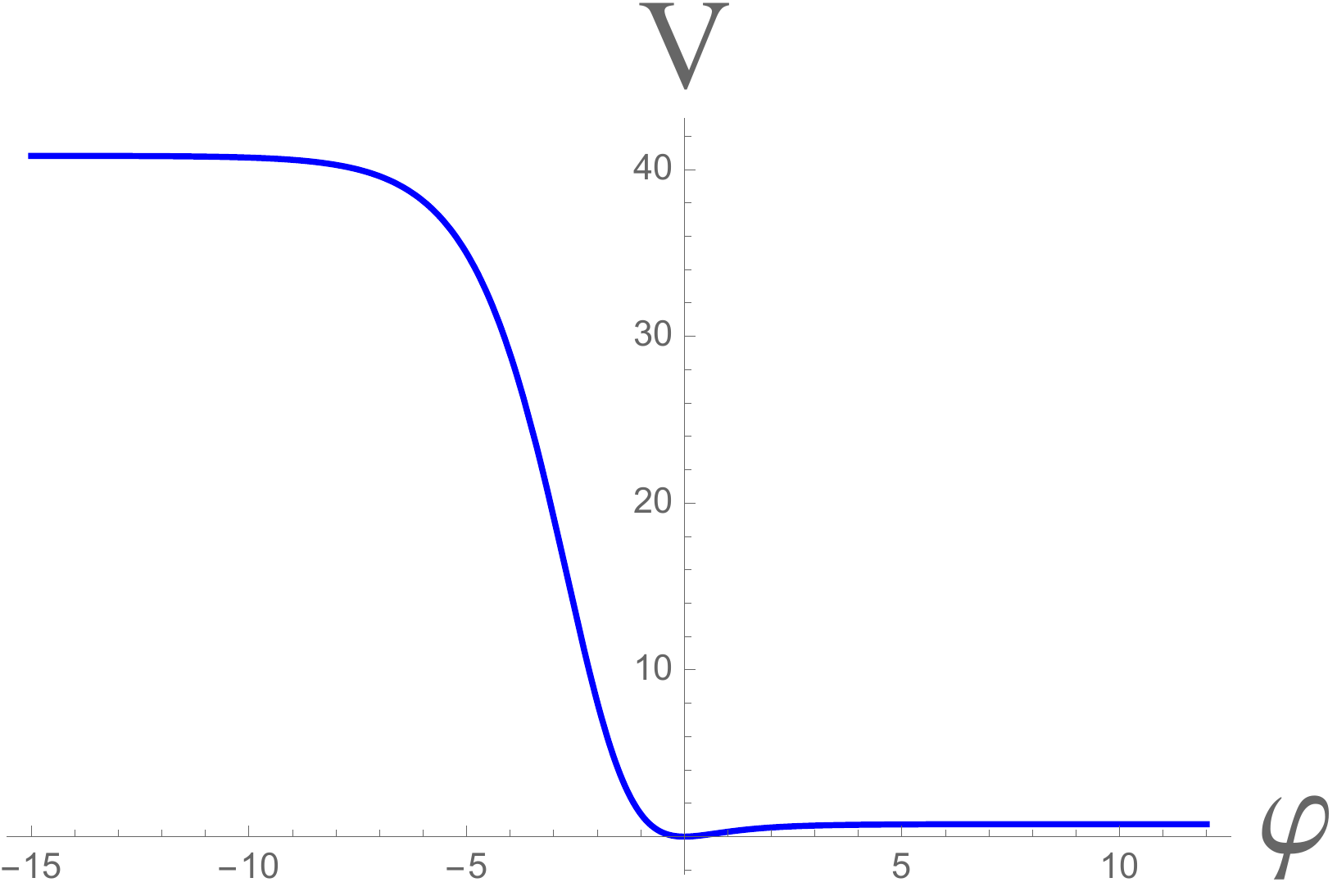}
}~~~~~
\caption{\footnotesize   Asymmetric shoulders.}\label{should}
\end{figure}

The situation becomes even more interesting in the theory
\be\label{expshoulders}
W^{\,\text{exp}}_{\mathbb{D}} =  \sqrt{\alpha}  \, \mu \, S \,  \, (1- e^{-\beta\,Z})\,  
\ee
with $\beta \gg 1$. Exponentially suppressed terms $e^{-\beta\,Z}$ may appear e.g. due to non-perturbative effects \cite{Kachru:2003aw}.  The potential in this case has a minimum at $Z = \vp = 0$. 
It has two shoulders of different height, each of which is capable of supporting inflation, with the inflaton potential
\be
V =  \alpha  \, \mu^{2} \, (1-e^{-\beta\, \tanh^{2n}{\vp\over \sqrt {6\alpha}}})^{2}\, .
\ee
Independently of $\alpha$, for $\beta\gg 1$, the two shoulders have relative height
${V_{+}/V_{-}} \approx e^{-2\beta}$. Thus for large $\beta$ the potential has two shoulders of exponentially different heights,  see Fig.~\ref{should} for a particular case of a potential with $\alpha = 1$, $\beta = 2$.  

 Despite the exponential difference in the height the values of the inflationary parameters $n_{s}$ and $r$ for inflation at each of these two shoulders coincide and are given by \rf{nsr}, at least for not too large values of $\alpha$. However, the amplitudes of scalar perturbations produced at these two branches are exponentially different from each other. At the upper branch the amplitude of the perturbations is proportional to $\mu (e^{\beta}-1)$, and at the lower branch it is proportional to $\mu (1-e^{- \beta})$, thus smaller by a factor of $e^{-\beta}$.

Suppose that the last stage of inflation, which determines the large scale structure of our universe, occurs at the lower shoulder with large positive $\vp$. Then, according to \cite{Kallosh:2015lwa}, one should take 
\be
V_{+}/\alpha \approx \mu^{2} \approx 10^{{-10}}
\ee
in Planck units. Considering $\alpha = O(1)$, one finds $V_{+} \sim 10^{{-10}}$. However, the left shoulder can be arbitrarily high, e.g,
by taking $\beta \sim 11$ one can easily have $V_{-}  = O(1)$ in Planck units in this model. This fact will be important for the discussion of the problem of initial conditions in this scenario. But before discussing the cosmological evolution in these models, we will return to the simplest symmetric T-models and study their potentials more attentively.

\section{Hyperbolic geometry and properties of T-model potential}\label{hyperbolic}
Consider the simplest T-model \rf{simpleT}, \rf{Vsimple}, see Fig.~1. We would like to describe this model in a more detailed way, which should help us to analyze initial conditions for inflation and the probability that it will take place in this model and its generalizations.  In order to do so, we will  the inflationary potential in a form most suitable for our investigation.

\begin{figure}[h!]
\centering
\includegraphics[scale=.28]{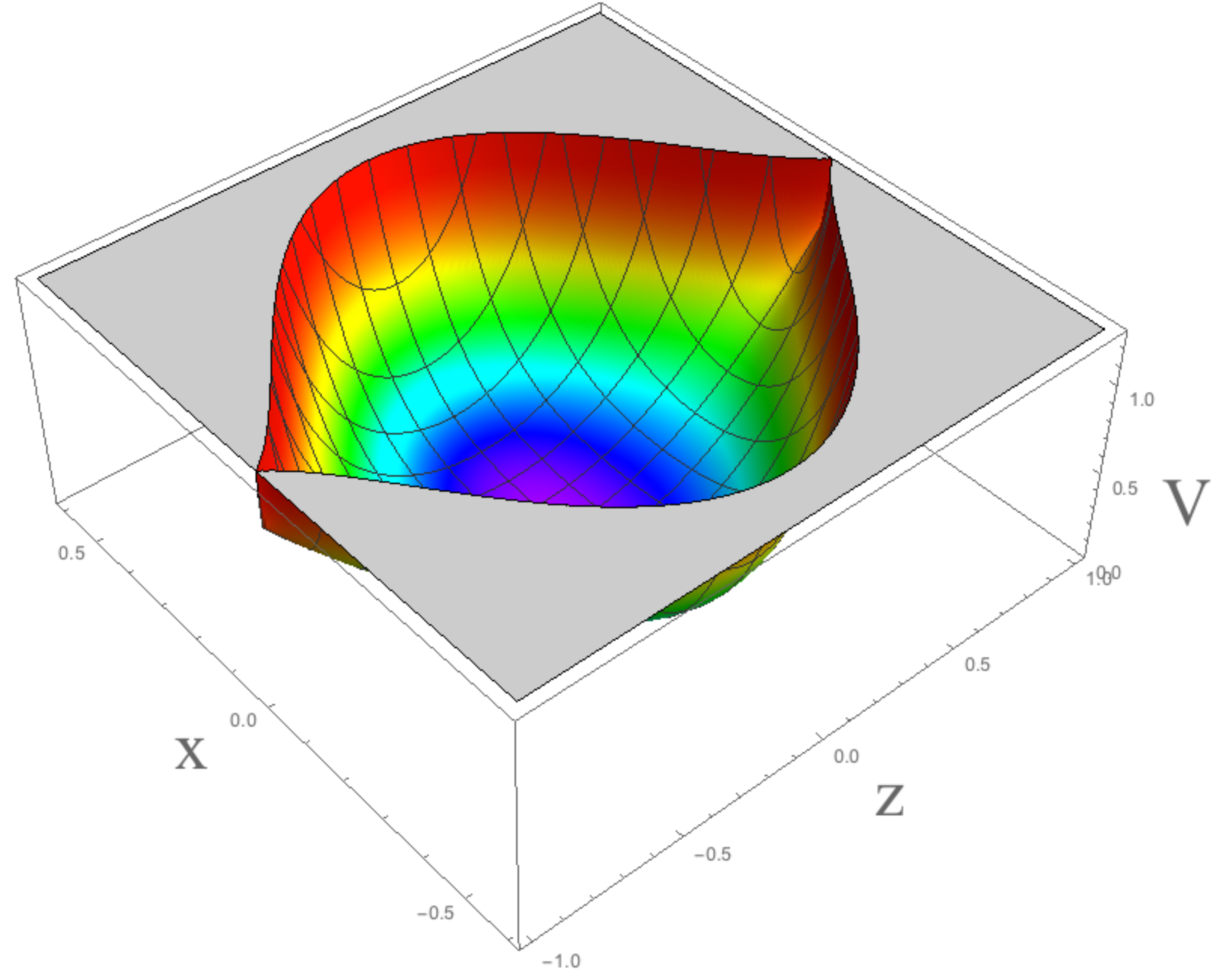}
\caption{\footnotesize Inflation, in terms of the variables $z$ and $x$, takes place in the two corners of the potential. The existence of the inflationary flat directions for $|z|-1 \ll 1$ is not apparent in these variables.}
\label{potz}
\end{figure}

In terms of $Z= z+ix$ the metric of the moduli space, which determines kinetic terms, and the potential are given by
\be
ds^2 =  {3\alpha\over (1- Z\bar Z)^2} d Z d \bar Z= 3\alpha {dz^2+dx^2\over (1-z^2-x^2)^2} \ ,
\ee
\be\label{orig}
V= \alpha\mu^{2}\ (z^2+x^2)\Big[{z^4+ 2z^2(x^2-1) + (x^2+1)^2\over (1-z^2-x^2)^2}\Big]^{3\alpha\over 2}  .
\ee
The existence of the flat inflaton direction is not obvious if one is looking at the potential \rf{orig} in the original variables $z$ and $x$, see Fig.~\ref{potz}. That is why one should try to represent the potential in terms of more adequate variables.

\begin{figure}[h!]
\centering
\includegraphics[scale=.27]{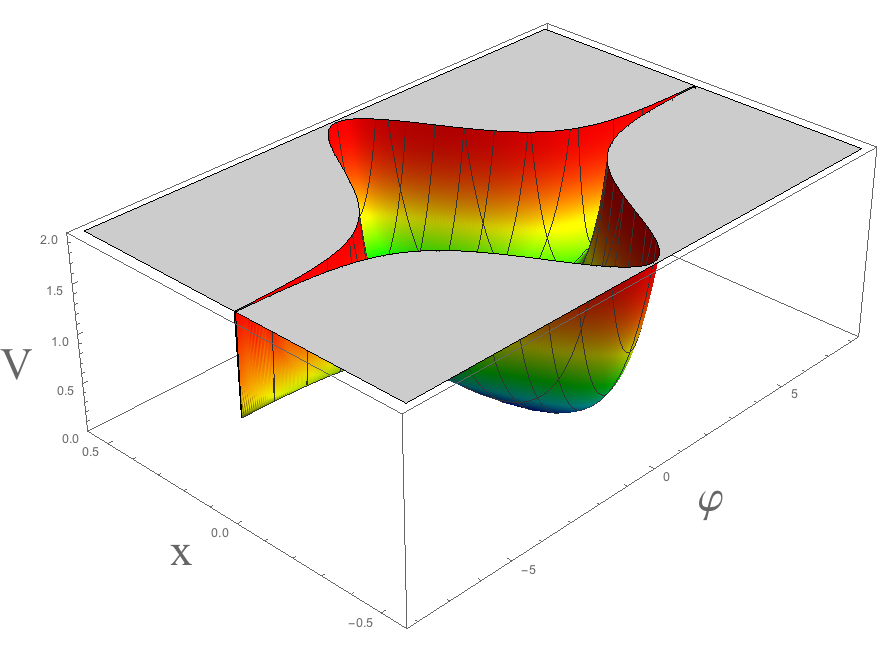}
\caption{\footnotesize A narrowing  trench in the potential in terms of $\vp$ and $x$ variables.  We see here the same effect as a decreasing size of angels and devils towards the boundary of the Poincar\'e disk in Escher's picture Circle Limit IV. However, this narrowing of the trench is just an illusion, which disappears when one plots the potential in proper coordinates, as shown in Fig.~\ref{tr}. }
\label{gorge}
\end{figure}

One can understand the situation better by making a change of variables $z = \tanh {\vp\over \sqrt {6\alpha}}$. For $x = 0$, the field $\vp$ plays a role of a canonically normalized inflaton field, and the existence of the inflationary shoulders of the
 potential becomes manifest in the variables $\vp$, $x$, see Fig.~\ref{gorge}. 
 
By looking at Fig.~\ref{gorge} one could get a wrong impression that the potential in the vicinity of the inflationary trajectory at $x = 0$ is incredibly steep: it looks like a gorge which becomes more and more narrow at large $|\vp|$.  However, this is just an illusion. As we have found, the curvature of this potential in the direction orthogonal to the inflationary trajectory is given by $2V(\vp) = 6H^{2}$, which is almost exactly constant during inflation, see \rf{mass}. We found this result by taking into account that the field $x$ is not canonically normalized. Now we will make this conclusion manifest by plotting the potential in terms of more adequate variables.

We will use the $\Phi$  variables 
as shown in eq. \rf{Phi} with the  kinetic terms in \rf{JJ}
The potential in these variables is
\be
V= \alpha\mu^{2}\  \left |\tanh
 {\vp + i \vt \over \sqrt {6\alpha}}\right|^{2}
\cdot  \Big (\cos \sqrt{2  \over 3\alpha} \vartheta   \Big ) ^{-{3\alpha}} ,
\label{Pot}\ee
where $\big|\tanh
 {\vp + i \vt \over \sqrt {6\alpha}}\big|^{2} = \tanh
{\vp + i \vt \over \sqrt {6\alpha}}
\cdot  \tanh
{\vp - i \vt \over \sqrt {6\alpha}}$. We may also present it in the form
\be
V= \alpha\mu^{2}\  {\cosh  \sqrt{2  \over 3\alpha} \vp - \cos  \sqrt{2  \over 3\alpha} \vartheta   \over \cosh  \sqrt{2  \over 3\alpha} \vp + \cos  \sqrt{2  \over 3\alpha} \vartheta  }
\cdot  \Big (\cos \sqrt{2  \over 3\alpha} \vartheta   \Big ) ^{-{3\alpha}} .
\label{Pot1}\ee

\begin{figure}[h!]
\centering
\includegraphics[scale=.29]{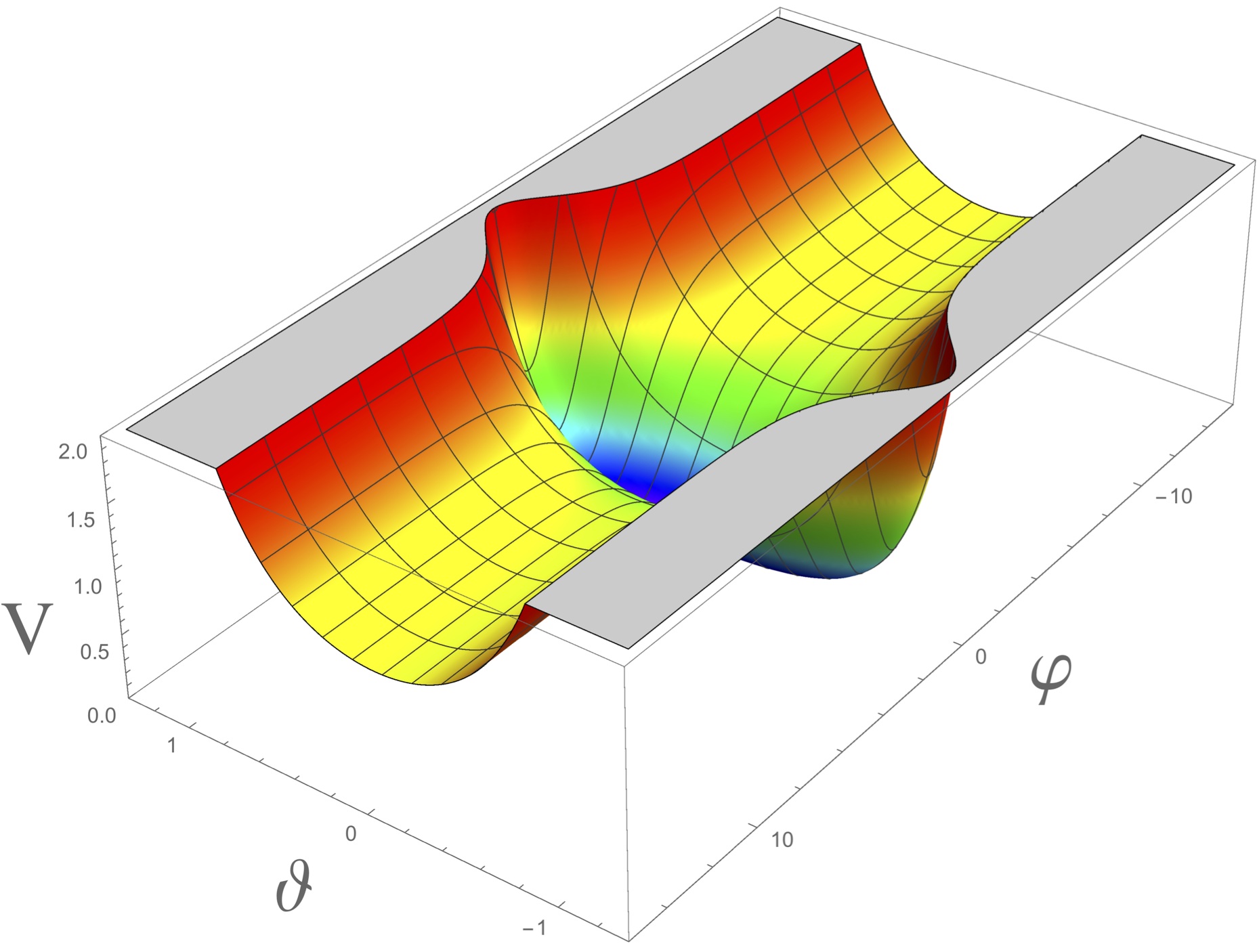}
\caption{\footnotesize  The T-model potential in terms of the variables $\vp$ and $\vt$ has two infinitely long dS valleys of constant width.}
\label{tr}
\end{figure} 

This potential is shown in Fig.~\ref{tr}.
It has a minimum at $\vartheta=0$ where the kinetic terms of both fields become canonical, $ds^2 \rightarrow    {1\over 2}  (d\vp^2 + d\vartheta^2)$ at $\vt \rightarrow 0$.
  At large values of $\vp$ where  $\tanh
 {\vp \over \sqrt {6\alpha}}$  approaches 1, the plot of the potential in terms of $\vp$ and $\vt$ has a dS valley of constant, $\vp$-independent width, instead of the rapidly narrowing gorge shown in Fig.~\ref{gorge}.
This fact will be very important for us shortly, when we will study the cosmological evolution of the fields $\vp$ and $\vt$ and initial conditions for inflation in these models. 

\begin{figure}[ht!]
\centering
\includegraphics[scale=.38]{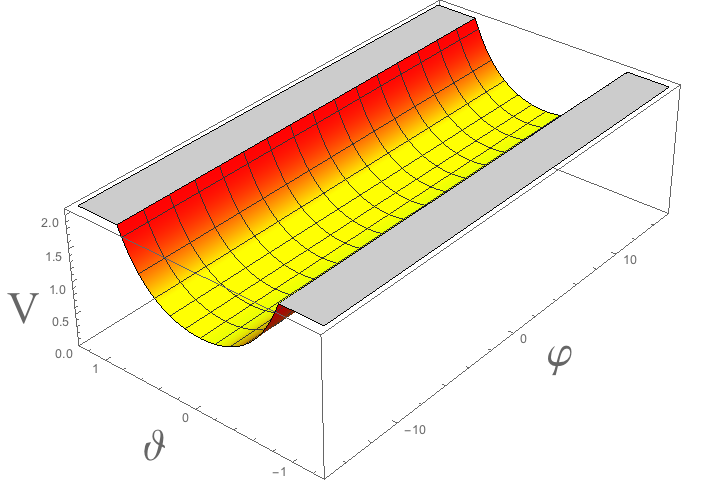}
\caption{\footnotesize An infinite $\vp$-independent dS valley \rf{PotdS} in the theory \rf{simpleT2} in $\vp$, $\vt$ coordinates. }
\label{gorgeexp2}
\end{figure}

For a better understanding of the structure of this potential, it is instructive to simplify even a little further the superpotential of our simplest T-model: Instead of $W = \sqrt{\alpha}  \, \mu \, S \,  \, Z\,$ \rf{simpleT}, let us consider a $Z$-independent superpotential
\be\label{simpleT2}
W =  \sqrt{\alpha}  \, \mu \, S   \ .
\ee
The potential in this model in the $\vp$ and $\vt$ variables is
\be
V= \alpha\mu^{2}\  \Big (\cos \sqrt{2  \over 3\alpha} \vt   \Big ) ^{-{3\alpha}}  .
\label{PotdS}\ee
Note that this potential does not depend on the inflaton field $\vp$, and has a dS minimum $V= \alpha\mu^{2}$ at $\vt = 0$. It represents an infinite $\vp$-independent dS valley as shown in Fig.~\ref{gorgeexp2}. 

As one can easily check,  the shape of this valley coincides with the shape of the dS valley in the simplest T-model \rf{Pot1} in the large $\vp$ limit.
This potential is manifestly shift-symmetric with respect to the field $\vp$. It is singular at $\cos \sqrt{2  \over 3\alpha} \vt \to 0$, but this singularity disappears if one uses canonical variables $\chi$ defined by
$
d\chi =  {d\vartheta \over \cos\sqrt{2  \over 3\alpha}}.
$
In the limit $\cos\sqrt{2  \over 3\alpha} \ll 1$,  which corresponds to $V \gg  \alpha\mu^{2}$, the  potential of the field $\vt$ in terms of the canonically normalized field $\chi$ acquires the following simple form:
\be\label{expx}
V(\chi) = \alpha\mu^{2} \left({3\alpha\over 2}\right)^{3 a/2}  e^{\sqrt{6 \alpha}\, \chi} \ .
\ee
For $\alpha \gtrsim 1$, this potential grows exponentially fast, and therefore the field $\chi$ should rapidly fall to the minimum of the potential and stay there. Meanwhile the situation for $\alpha \ll  1/3$ is quite different. It is well known that the potentials of that type are sufficiently flat to support inflation \cite{Liddle:1988tb}. Indeed, in the theory \rf{expx} the universe expands as $t^{1/3\alpha}$. The defining condition for inflation is $\dot H \ll H^{2}$, which means that the Hubble parameter (and the size of the horizon) does not change much during the Hubble time $H^{{-1}}$. This condition is satisfied for $\alpha \ll 1/3$.  Thus  for $\alpha \ll 1/3$  inflation can begin at very large values of $V(\vp,\vt)$ corresponding to large values of $\chi$, and then the universe inflates while the field $\chi$ slowly rolls down to the bottom of the dS valley. After that, inflation becomes driven by the field $\vp$.

\begin{figure}[ht!]
\centering
\includegraphics[scale=.34]{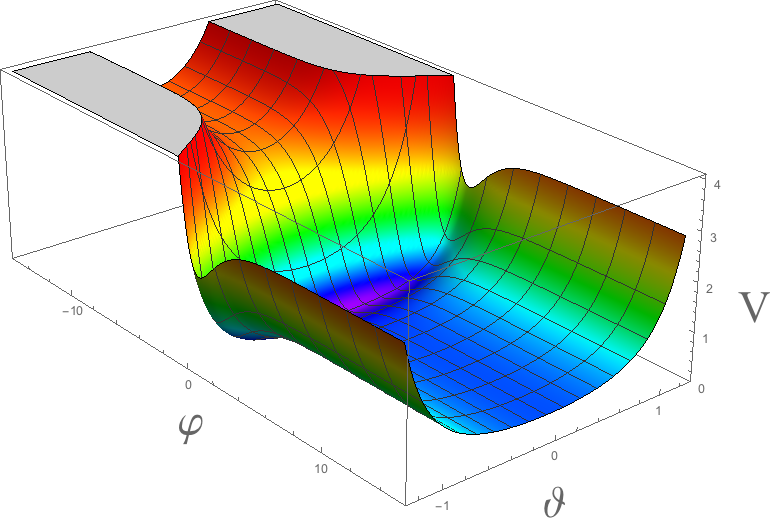}
\caption{\footnotesize Two-shoulder potential in the theory \rf{expshoulders} in $\vp$, $\vt$ coordinates. }
\label{gorgeexp}
\end{figure}

The existence of the dS valley of a constant, $\vp$-independent width will play a crucial role in our  discussion of initial conditions for inflation. Therefore we would like to emphasize that the existence of such dS valleys  is not just a property of the simplest T-model \rf{simpleT}, \rf{Pot1}, but a general property of the class of models \rf{NewD}, \rf{gen} studied in this paper. In particular, the potential of the asymmetric T-models \rf{lin}, \rf{expshoulders} exhibits two infinite dS valleys of constant width but different height, as shown in Fig.~\ref{gorgeexp}. Meanwhile in E-models, there is one infinite dS valley at $\vp > 0$.

In the next sections, we will describe the cosmological evolution in the simplest T-model with potential \rf{Pot1} for $\alpha > 1/3$, as well as for $\alpha \ll 1/3$.

\section{Other examples of $\alpha$-attractors}\label{aaa}

Before moving further to the discussion of the cosmological evolution in this class of models, we will consider another class of $\alpha$-attractors, with a slightly different \K\ potential as compared to \rf{JJ1}, with $S\bar S$ placed under the sign of the logarithm:
\be
 K= -3\alpha \log \Big [\cosh \Big ({\Phi-\bar \Phi \over \sqrt{6\alpha}}\Big) - S \bar{S}\Big] \ .
\label{JJ12}\ee
We will consider the simplest superpotential $W   = \sqrt{\alpha} \,  \mu \, S \, Z  = \sqrt{\alpha} \,  \mu   \tanh
 {\Phi \over \sqrt {6\alpha}}$. In this case, the scalar potential is
\be
V= \alpha\mu^{2}\  {\cosh  \sqrt{2  \over 3\alpha} \vp - \cos  \sqrt{2  \over 3\alpha} \vartheta   \over \cosh  \sqrt{2  \over 3\alpha} \vp + \cos  \sqrt{2  \over 3\alpha} \vartheta  }
\cdot  \Big (\cos \sqrt{2  \over 3\alpha} \vartheta   \Big ) ^{1-{3\alpha}} .
\label{Pot2}\ee
Note that this potential almost exactly coincides with \rf{Pot1}. The only difference is that instead of the term $ \big(\cos \sqrt{2  \over 3\alpha} \vartheta   \big) ^{-{3\alpha}} $  in \rf{Pot1} now we have $ \big(\cos \sqrt{2  \over 3\alpha} \vartheta  \big) ^{1-{3\alpha}}$.  This leads to the stability condition $\alpha \geq 1/3$. For $\alpha > 1/3$ the potential has a dS valley very similar to the one shown in Fig. \ref{tr}, with curvature  given by $m^{2}_{\vt} = 6H^{2}(1-{1\over 3\alpha})$. For the special case $\alpha = 1/3$, the potential looks like a disk \cite{Kallosh:2015zsa}, see Fig. \ref{gorgeexp3}
 in Section \ref{generalizations}. 

\section{Inflation in a homogeneous universe} 

In this section we will study the evolution of the fields $\vp$ and $\vt$ during inflation in the FRW background with various initial conditions. Before analyzing this evolution numerically, we first study it analytically in simplified cases, which will help us to build intuition instead of fully relying on results of numerical investigation.

Consider the evolution of the field $\vp$ along the inflationary dS valley at $\vt = 0$ in the limiting case $\vp/\sqrt\alpha \gg 1$, when the potential in the $\vp $ direction is constant, $V_{dS} = \alpha\mu^{2}$, with exponentially good accuracy. Suppose that $V\ll 1$ but the initial velocity of the field $\vp$ is extremely large, so that its kinetic energy is much greater than the potential energy, $\dot\vp^{2}/2 \gg V_{dS} =\alpha\mu^{2}$. Note that $\mu^{2} \approx10^{-10}$ in the models describing inflationary spectra with Planck 2015 normalization \cite{Kallosh:2015lwa}, so for $\alpha = O(1)$ one should have $V_{dS} \sim 10^{-10}$.  How far will the field $\vp$  travel from its initial value $\phi_{0}$ in a space with energy density $\vp^{2}/2 + V_{dS}$, if initially the universe was in the Planck energy state dominated by kinetic energy of the field $\dot\vp^{2}/2 = 1$?

The equation of motion of the canonical field $\vp$ at $\vt = 0$ is $\ddot\vp + 3 H \dot\vp = 0$, which in our case satisfies:
\be
\label{dot}
\ddot\vp + \sqrt{3} \dot\vp \sqrt{\dot\vp^{2}/2 +V} =0 .
\ee
This is simple enough so as to allow a full analytical solution, but instead we will solve it approximately, which will be helpful for understanding the general situation.

We will start with initial conditions such that $\dot\vp^{2}/2 = N_{k}\, V$, where $N_{k}$ is some number much greater than 1. If initially the kinetic energy of the scalar field was Planckian and $V_{dS} \sim 10^{{-10}}$, we have $N_{k} \sim 10^{{10}}$. Then, during the first stage of the field evolution, one can entirely ignore its potential energy, so that the equation becomes $\ddot\vp =- \sqrt{3/ 2}~\dot\vp^{2}$, with the solution \cite{Felder:2002jk} 
\be
\label{dot2}
{\dot\vp^{2}\over 2} ={1\over 3t^{2}} , \qquad \vp(t) - \vp_{0} = \sqrt{2\over 3} \, \ln{t\over t_{0}} .
\ee
This kinetic energy dominated regime continues from the initial state with ${\dot\vp^{2}\over 2} = N_{k} V_{dS}$ to the point where it becomes approximately equal to $V_{dS}$. During this time, the field $\vp$ moves by $\Delta\vp = {1\over \sqrt 6} \ln N_{k} \sim 9.4$ in Planck units, for $N_{k} \sim 10^{10}$.

After that, the kinetic energy of the field becomes sub-dominant, and it continues moving in accordance with equation $\ddot\vp =- \dot\vp \sqrt{3V_{dS}}$. One can show that it passes a distance slightly smaller than 1 and  stops. Thus the total shift of the field $\vp$ due to its initial velocity is about 10 Planck units, independently of $\alpha$, after which all memory about the initial velocity of the field at the Planckian time completely disappears. This conclusion is confirmed by the results of numerical calculations.

In this investigation, we ignored the motion of the field $\vt$. A combined motion of these two fields for various $\alpha$ is more difficult to analyze, but our numerical calculations show that the final result remains the same: If one starts with general initial conditions corresponding to the Planck energy density, including kinetic and potential energy of all fields, and follow the evolution of these fields at large $\vp$ where the potential is shift-symmetric with respect to $\vp$, then the maximal distance the field $\vp$ may travel from its initial position until it enters the slow-roll inflationary regime is $\Delta\vp = {1\over \sqrt 6} \ln N_{k} \sim 9.4$.

In the slow-roll inflationary regime, which starts when nearly all of the initial kinetic energy of the field $\vp$ is lost, this field $\vp$ rolls towards the minimum of the potential by the distance $\delta\vp = \sqrt{3\alpha\over 2}\, \ln{8N_{e}\over 3}$ until inflation ends \cite{Kallosh:2013hoa,Kallosh:2013yoa}. Here $N_{e}$ is the number of e-foldings during inflation. For example, for $\alpha = 1 $ and $N_{e} =60$ one has $\delta\vp \approx 6$. 

Thus we are coming to a rather interesting conclusion: Even if the inflationary plateau is extremely low, compared to the Planck density, $V_{dS}\sim 10^{-10}$,  initial motion of the fields $\vp$ and $\vt$ does not affect the onset of the inflationary regime, at this plateau, if inflation begins at a distance $\delta \vp \gtrsim 10$  from the initial value of the field $\vp_{0}$ at the Planck time. For the simple T-models with $\alpha = O(1)$, this means that the last 60 e-folds of inflation do not depend on the initial velocities and of the fields $\vp$ and $\vt$ if the initial value of the field $\vp$  was $|\vp_{0}| > \Delta \vp +\delta\vp \gtrsim 16$. This conclusion will be important for us later, when we will discuss a more general situation where the universe in the beginning of its evolution may be very inhomogeneous.

\begin{figure}[h!]
\centering
\includegraphics[scale=.4]{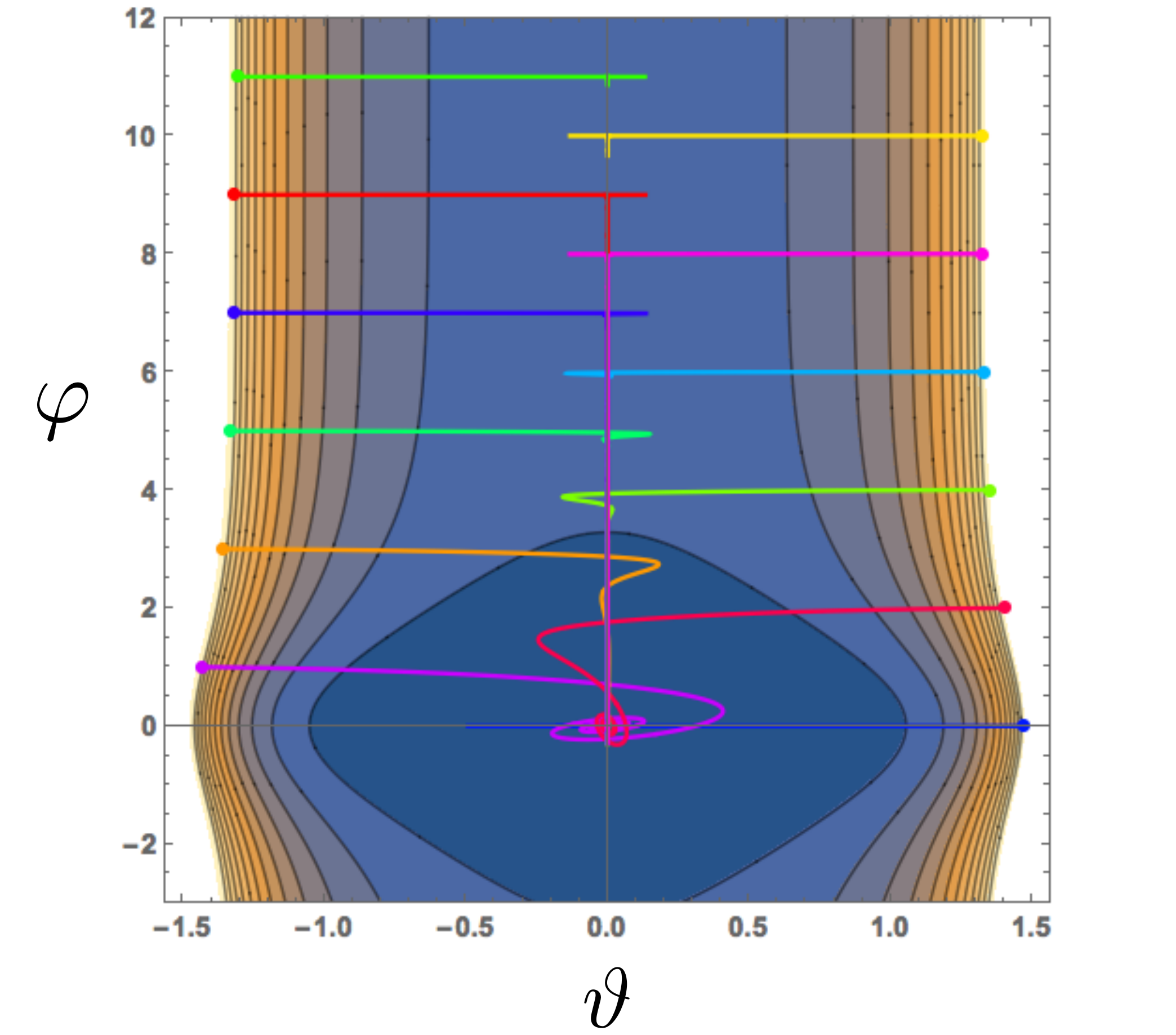}
\caption{\footnotesize Time evolution of scalars on a contour plot of the potential. In the beginning, the field $\vt$ moves towards the minimum of the dS valley, whereas the field $\vp$ remains nearly constant, if its initial value was large. Then after a short stage of oscillations, the  field $\vt$ vanishes, and the cosmological expansion becomes determined only by the inflaton field $\vp$, which slowly moves towards $\vp = 0$ and oscillates there. }
\label{run}
\end{figure}

To illustrate this conclusion, we present here some of the results of our numerical calculations for a particular case $\alpha = 1$, using a code developed in \cite{Kallosh:2004rs}. Using the kinetic term in \rf{JJ}, and the potential in \rf{Pot1} given as a function of $\vp$ and $\vt$, we solve the Friedmann equations for the time evolution of scalars for a set of initial conditions. We  plot the trajectories on a contour plot of the potential,  the values of the fields as a function of time and the scale factor  as a function of time. 
The examples here involve various choices of initial positions, as shown in Fig.~\ref{run} and vanishing initial velocities. 


To develop some intuition, we begin with the case with $\alpha = 1$ where the scalar fields start with zero initial velocities at the height 10 times above the bottom of the dS valley with $V_{dS} = \alpha\mu^{2}$. This will show us what the fields 'want' to do if not pushed hard, and whether this may lead to inflationary regime. 

\begin{figure}[h!]
\centering
\includegraphics[scale=.39]{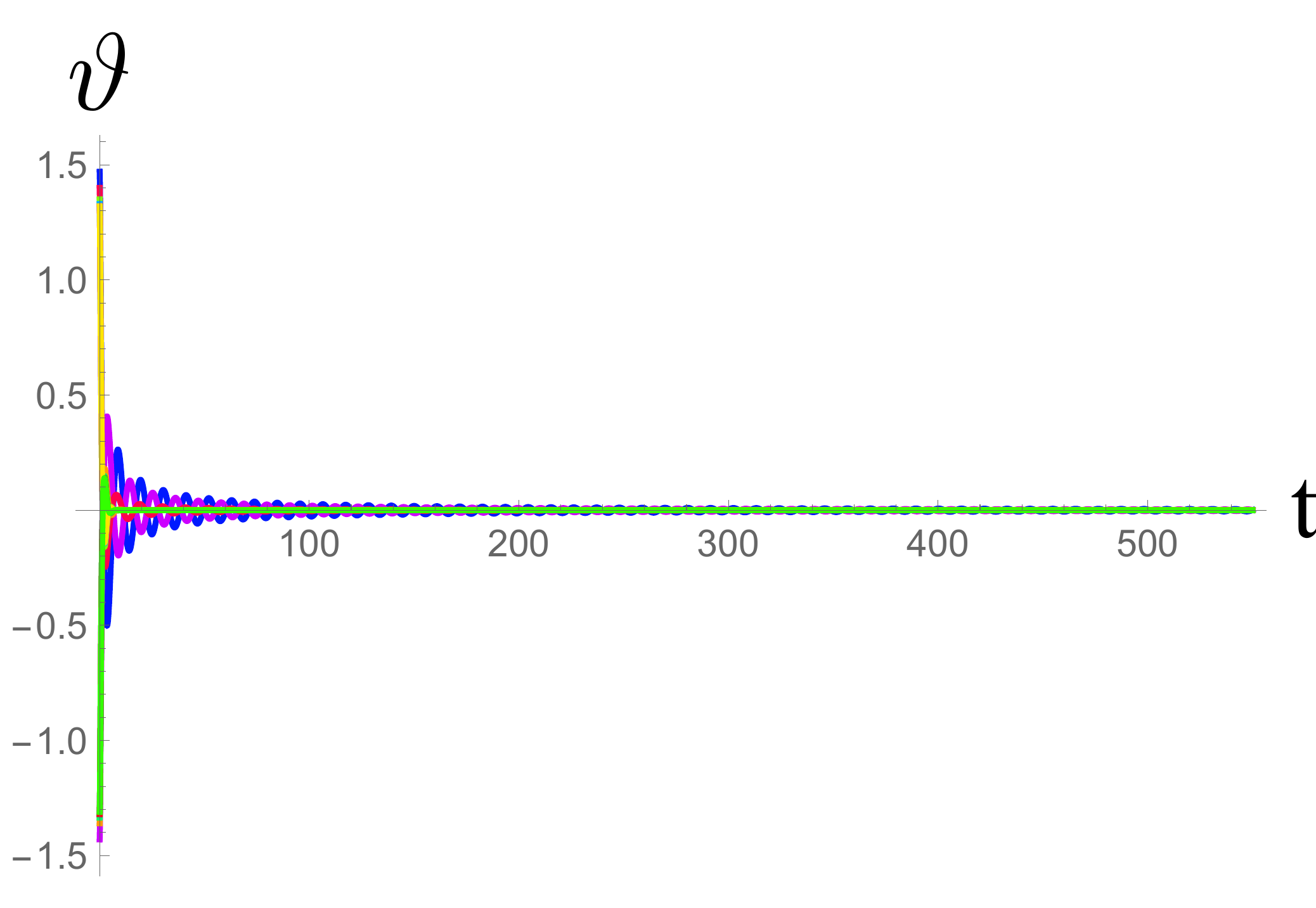}
\caption{\footnotesize  The field $\vt$ moves quickly towards the minimum, at the bottom of the dS valley. Then after a short stage of oscillations, the  field $\vt$ vanishes and remains in all cases at the bottom of the dS valley. }
\label{EvolutionTheta}
\end{figure}

\begin{figure}[h!]
\centering
\includegraphics[scale=.32]{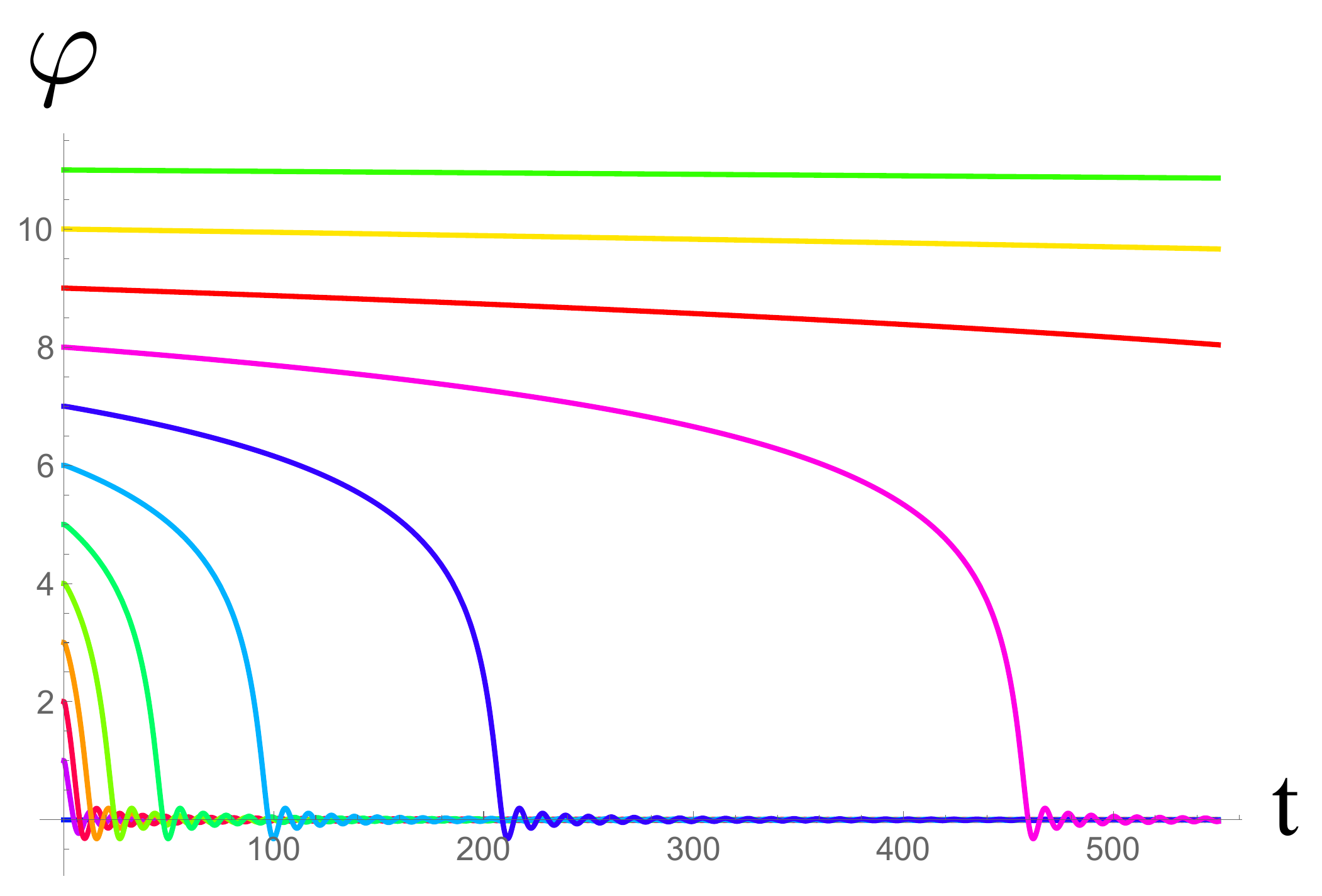}
\caption{\footnotesize The field $\vp$ remains nearly constant and very slowly moves towards $\vp = 0$ and oscillates there. The cases with smaller initial values of $\vp< 6$ reach the minimum earlier and inflate less. }
\label{EvolutionPhi}
\end{figure}

\begin{figure}[h!]
\centering
\includegraphics[scale=.35]{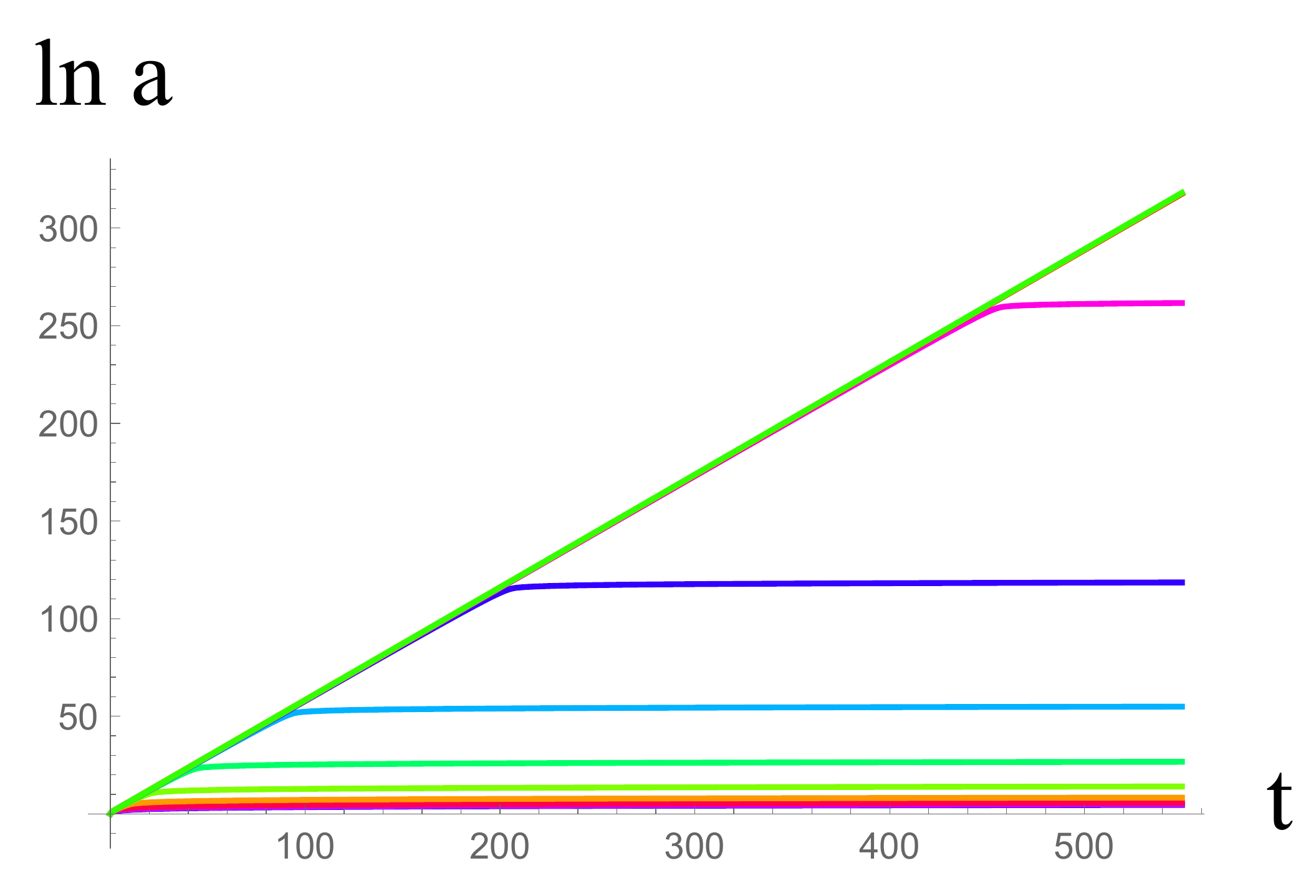}
\caption{\footnotesize  The $\log$ of a scale factor is plotted, as a function of time. The models with smallest, in our examples,  initial values of $\vp<6 $  produce less than 50 e-foldings, however, all other cases with $\vp> 6$ easily lead to more than 60. }
\label{EvolutionLoga}
\end{figure}

Figs. \ref{run}, \ref{EvolutionTheta}, \ref{EvolutionPhi} and \ref{EvolutionLoga} show 12 different trajectories with different initial values of the fields such that $V(\phi_{0},\theta_{0}) = 10 \alpha\mu^{2}$. The initial positions of the fields are shown by small circles. In all cases, the field $\vt$ rapidly rolls down to the bottom of the valley, oscillates and relaxes there. After that, for $\vp_{0} \gtrsim 3$ (in Planck units), the universe enters the stage of inflation driven by the field $\vp$. This stage is longer than 60 e-foldings for $\vp_{0} \gtrsim 6$.


Examples above and the corresponding Figures had same value of the initial energy density due to a proper choice of the initial values of $\vp_0$ and $\vt_0$ at $t=0$. However, the initial values of velocities were chosen to be zero, and the initial values of the potential energy was sub-Planckian. Here we would like to analyze a situation where the total value of the initial energy, kinetic and potential, is of a Planckian scale. Namely, here we will start with initial condition with velocities and positions such that at $a(0)=1$ the initial energy in Planck units $M_{p}=1$ is
\be\label{initial}
E_0= E_{kin}^0 + V_0 =1,
\ee
where
\be
E_{kin} = {\dot \vp^2 + \dot \vartheta^2 \over 2\cos^{2}\sqrt{2  \over 3\alpha} \vartheta} \ , \ee
and the potential is given by \rf{Pot1}. For the numerical calculations we took the height of the dS valley of the potential at $\vt=0$ and large $\vp$ as $\alpha \mu^{2} =10^{-10}$ in Planck units.

Our potential has a dS valley at a very small height $V_{dS} \sim 10^{-10}$ in Planck units, however, when $\vt$ is close to the value where $\cos\sqrt{2  \over 3\alpha} \vt$ vanishes,  the potential  \rf{Pot1} is extremely steep and easily reaches the Planck values $V = O(1)$.  A more proper way to understand the growth of the potential at large $\vt$ is to replace $\vt$ by its canonical counterpart $\chi$, as we did in \rf{expx}. The singular growth of potential close to the point where $\cos\sqrt{2  \over 3\alpha} \vt = 0$ is replaced by exponential growth at large $\chi$. For $\alpha \gtrsim 1/3$ the potential  \rf{expx} is very steep and field $\chi$ rapidly falls towards the dS valley. However, for $\alpha \ll 1/3$ the potential is sufficiently flat to allow inflation at all sufficiently large values of $V  \gg V_{dS} \sim 10^{-10}$.

\begin{figure}[h!]
\centering
\includegraphics[scale=.4]{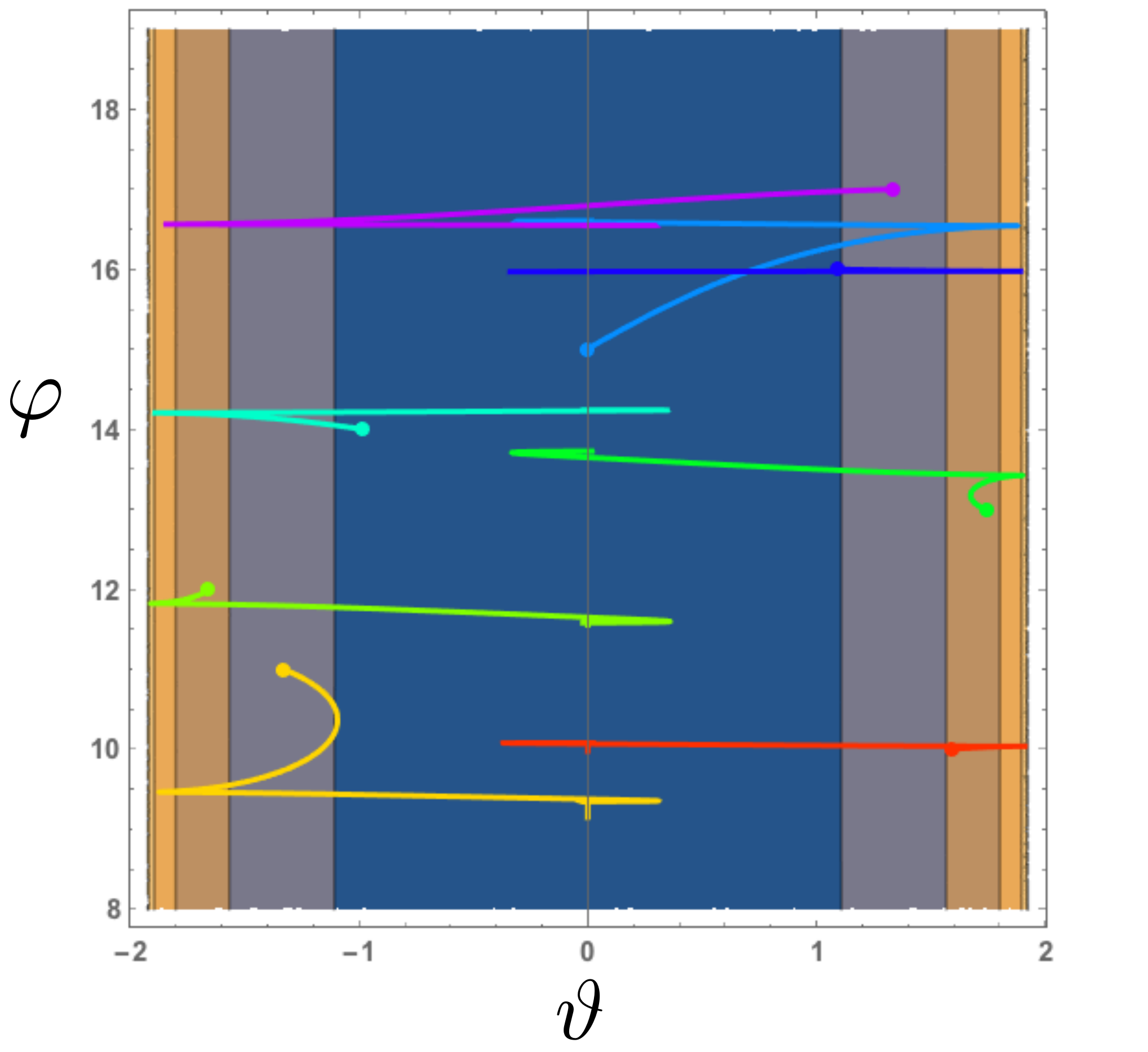}
\caption{\footnotesize  Evolution of the fields $\vp$ and $\vt$ starting at the Planck energy density.The fields start on the wall with near maximal $\vt$ or anywhere between the wall and the bottom of the dS valley with velocities such that the total energy is always $E_{kin}^0 + V_0 =1$.  In this process their total energy is decreasing to $\sim 10^{-10}$ when they reach the bottom of the valley. }
\label{FromPlanck}
\end{figure}

We start at Planckian energies $E_0$ by making random choices of $\vt,\vp, \dot \vt, \dot \vp$ such that Eq. \rf{initial} is satisfied and solve all equations numerically. Our goal is to find the trajectory of the fields starting from the Planckian total energy and find out how they reach the bottom of the dS valley loosing the total energy by 10 orders of magnitude. We show a sample of such trajectories on the contour plot of the potential in Fig.~\ref{FromPlanck}. For illustration purposes, we took all initial conditions at $\vp \gtrsim 9$ to illustrate what happens at sufficiently large $\vp$ along the infinitely long dS valley, where the potential practically does not depend on $\vp$.

\begin{figure}[h!]
\centering
\includegraphics[scale=.43]{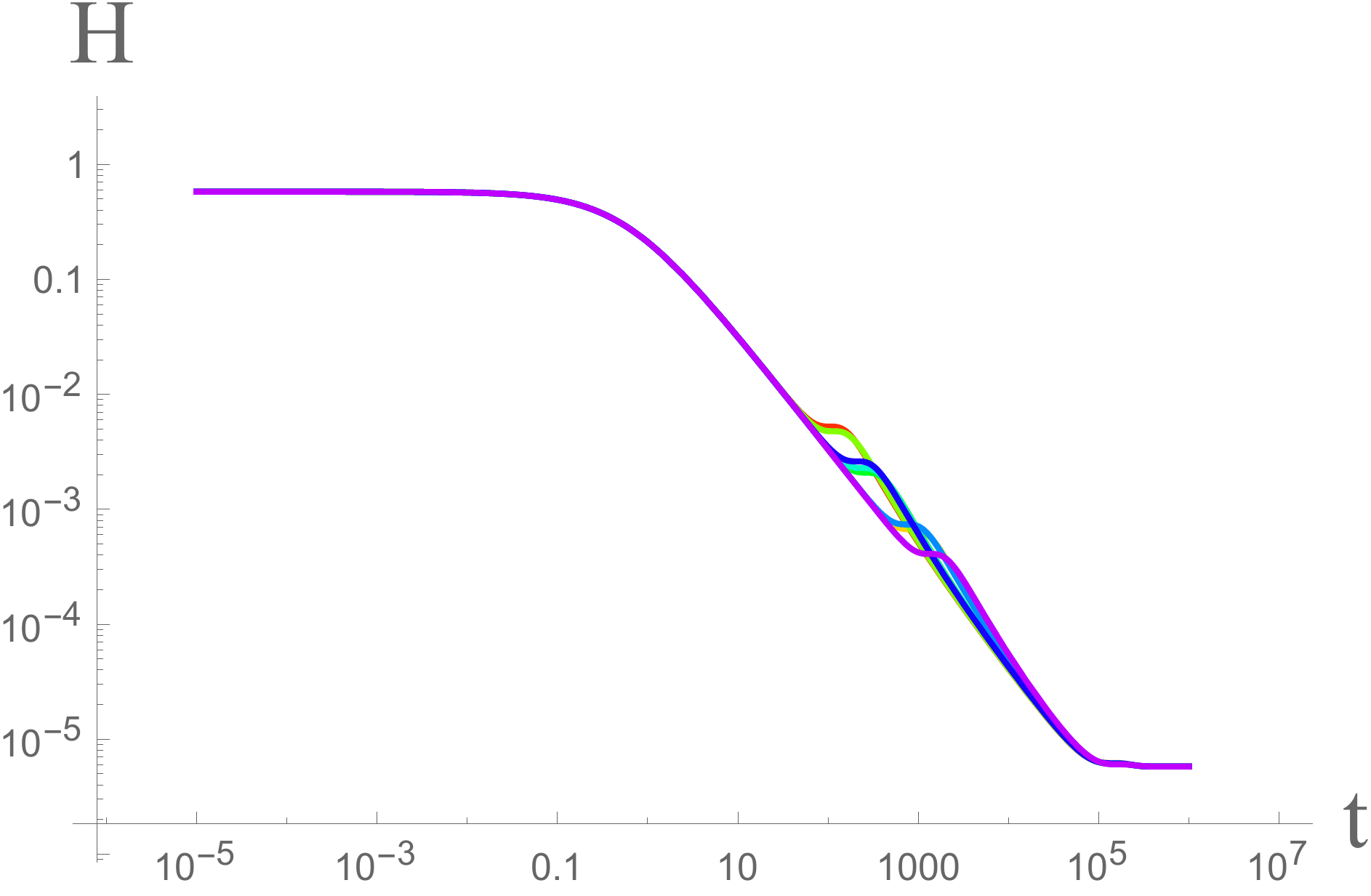}
\caption{\footnotesize  The Hubble parameter for all initial conditions in Fig.~\ref{FromPlanck} starts at $3H_0^2=1$, $H_0= {1\over \sqrt 3}$, and ends at the bottom of dS valley with $V \sim 10^{{-10}}$. The intermediate stage is somewhat different for examples with various positions and velocities. }
\label{HubblePlanck}
\end{figure} 

The fields start on the wall with near maximal $\vt$ or anywhere between the wall and the bottom of the dS valley with velocities such that the total energy is always $E_{kin}^0 + V_0 =1$.  In this process their total energy is decreasing to $V_{dS} \sim 10^{-10}$ when they reach the bottom of the valley. It may not be apparent from this figure, but after a series of oscillations with a rapidly decreasing amplitude, similar to what was shown in Fig.~\ref{EvolutionPhi}, the fields relax at the bottom of the valley. For large values of $\vp$, the potential along the valley is almost exactly flat. Therefore the inflaton field stay almost still for a long time until it start slowly rolling to the minimum of the potential at $\vp = 0$. The beginning of this slow motion is only barely seen here in the end of the lower (yellow) trajectory, which just starts to move to smaller values of $\vp$.

In Fig.~\ref{HubblePlanck} we show the time evolution of the Hubble parameter. For all initial conditions in Fig.~\ref{FromPlanck}, the Hubble constant  $H$ starts at $3H_0^2=E_0 =1$, i.e. $H_0= {1\over \sqrt 3}$, and ends at the bottom of dS valley with $V \sim 10^{{-10}}$. The intermediate stage is somewhat different for examples with various positions and velocities, however, eventually they all end up with $3 H^2 \approx V_{dS} \approx 10^{-10}$, $H\approx {1\over \sqrt 3} 10^{-5}$. The intermediate flattening of the Hubble parameter shown in the Figure has an interesting origin: it is due to $\vt$ reaching its maximum, before it goes down to zero.

As we see, if we start with sufficiently large initial value of the field $\vp_{0} \gtrsim 10$, then, independently of the choice of other initial conditions at the Planck density with $E_{kin}^0 + V_0 =1$, the fields fall to the bottom of the valley and slow down there in such a way that the field $\vp$ does not change much in this process, in complete agreement with our general expectations.  Thus, all choices of initial conditions with arbitrary velocities of the fields and arbitrary deviations of $\vt$ from zero (from the bottom of the valley) will lead to inflation if the initial value of the field $\vp$ took any value in the infinitely long interval $10\lesssim |\vp_{0}| < \infty$.  

So far, we studied the cosmological evolution in a homogeneous universe, with the same values of the fields everywhere. Now we are going to discuss a more general case where the universe may be non-uniform. Before doing it, we will give a brief review of the  problem of initial conditions for inflation in an inhomogeneous universe, and later on we will return to the cosmological attractors in this context.

\section{Initial conditions for inflation in an inhomogeneous universe: A brief review} \label{review}

The theory of initial conditions for inflation was developed simultaneously with the invention of the chaotic inflation scenario \cite{Linde:1983gd,Linde:1984ir,Vilenkin:1984wp,Linde:1985ub,Linde:1986fd,Linde:2005ht}, see \cite{Linde:2014nna} for a recent discussion. In the previous sections we discussed only the homogeneous universe case, where the situation is relatively simple. Now we are going to consider the initial condition problem in a more general situation, when the universe may be inhomogeneous.

Let us remember first what was the main problem with the hot Big Bang scenario in this respect. In that model, the universe was born at the cosmological singularity, but it was possible to describe it in terms of classical space-time only when time is greater than the Planck time $t_{p} \sim 1$. At that time,  the temperature of matter was given by the Planck temperature $T_{p} \sim 1$, and the density of the universe was given by the Planck density $\rho_{p} \sim  1$. The size of the causally connected part of the universe at the Planck time was $ct_{p} \sim 1$. Each such part contained a single particle with the Planck temperature. The subsequent evolution was supposed to be adiabatic.  As such, the total number of such particles is approximately conserved, so it must be greater than the total number of particles in the observable part of the universe, $n \sim 10^{{90}}$. This means that the universe at the Planck time consisted of $10^{90}$ causally disconnected parts. The probability that all of these independent parts emerged from singularity at the same time with the same energy density and pressure is smaller than $e^{-10^{90}}$, and the probability that our universe from the very beginning was uniform with an accuracy better than $10^{-4}$ is even much smaller.

The change of the perspective brought by the invention of the chaotic inflation scenario was quite dramatic. The main condition required in the simplest models of chaotic inflation was the existence of a single Planck size domain where the kinetic and gradient energy of the scalar field is few times smaller than its potential energy $V(\phi) \sim 1$. For sufficiently flat potentials, it leads to inflation, so the whole universe appears as a result of expansion of a single Planck size domain. According to  \cite{Linde:1983gd,Linde:1984ir,Vilenkin:1984wp,Linde:1985ub,Linde:1986fd,Linde:2005ht}, the probability of this process is not exponentially suppressed. After that, the universe enters an infinite process of self-reproduction \cite{Linde:1986fd}.

However, in the argument given above it was assumed that inflation may start at the densities comparable with the Planck density. This condition is not satisfied in hilltop models \cite{Linde:1981mu}, in the Starobinsky model \cite{Starobinsky:1980te}, in the Higgs inflation model  \cite{Salopek:1988qh}, and in the simplest versions of the broad class of the cosmological attractor models discussed in this paper and in \cite{Kallosh:2013tua,Galante:2014ifa}, where inflation occurs only at $V\ll 1$. Thus the issue of initial conditions in these models should be addressed. 

To explain the main problem, we will consider the new inflation/hilltop models  \cite{Linde:1981mu} and follow the argument against them and in favor of chaotic inflation given in  \cite{Linde:2005ht}. Inflation in such models begins at the density about 10 orders of magnitude below the Planck density. If the universe is closed, then a typical closed universe born at the Planck density and filled with relativistic particles collapses within the Planck time, unless it is extremely large from the very beginning  \cite{Linde:2005ht}. But then inflation does not explain why our universe is large. Rather, inflation in such models {\em requires} the universe to be large. 

Moreover,  it is hard to expect a large universe to be uniform on the scale $H^{{-1}}$, as required for inflation, at the time when the energy density of normal matter drops down to the energy density at the top of the inflationary potential $V \sim 10^{{-10}}$. 
Indeed, the size of the Hubble domain at that time is about $t\sim 10^{5}$, but the total size of the hot universe was growing only as $\sqrt t$. As a result, the Hubble size domain at the moment when inflation could start contains $t^{3/2} \sim 10^7$ parts which had the Planck size and were causally disconnected at the Planck time. The probability that the universe in all of such domains had the same density can be estimated by $e^{-10^{7}}$. It is much better than $e^{-10^{90}}$, but it is still highly unsatisfactory \cite{Linde:2005ht}. Recently, a very similar argument against naturalness of initial conditions in the Starobinsky model was given in \cite{Dalianis:2015fpa}. Does it mean that all such models are in trouble?

Fortunately, several ways to solve the problem of initial conditions in the models with low scale of inflation have been proposed more than 10 years ago, see a discussion of this issue in  \cite{Linde:2014nna} and references therein. One of the simplest solutions is to consider a flat compact universe having the topology of a
torus, $S_1^3$,
\begin{equation}\label{2t}
ds^2 = dt^2 -a_i^2(t)\,dx_i^2
\end{equation}
with identification $x_i+1 = x_i$ for each of the three dimensions \cite{Linde:2004nz}. Whereas this may sound exotic, such models are routinely used by string theorists who assume that 6 internal dimensions are compactified, and by experts in numerical relativity who study the universe on a lattice with periodic boundary conditions. The trick is to make this box exponentially big and its periodicity unobservable, which can be easily done by inflation.

Suppose,
for  simplicity, 
that $a_1 = a_2 = a_3 = a(t)$. In this case the curvature
of the universe and the Einstein equations in terms of $a(t)$ will be
the same as in an infinite flat Friedmann universe with the metric $ds^2 = dt^2
-a^2(t)\,d{\bf x^2}$. In our notation, the scale factor $a(t)$ is equal to the
size of the universe in Planck units.

Let us assume that at the
Planck time $t_p \sim 1$ the universe was radiation dominated, $V\ll
 T^4 = O(1)$ and that at the Planck time the total size of the
box was Planckian, $a(t_p) = O(1)$. Then the whole universe initially
contained only $O(1)$ relativistic particles such as photons or gravitons, so
that the total entropy of the whole universe was $O(1)$.

 The size of the universe dominated by relativistic particles  grows as
$a(t) \sim \sqrt t$, whereas the mean free path of the gravitons grows as
$H^{-1}\sim t$. If the initial size of the universe was $O(1)$, then at the
time  $t \gg 1$ each particle (or gravitational perturbation of the metric)
within one cosmological time would run all over the torus many times, appearing
in all of its parts with nearly equal probability. This effect, called
``chaotic mixing,'' should lead to a rapid homogenization of the universe
\cite{Linde:2004nz,chaotmix,topol4}. 

In fact, to achieve a modest degree of homogeneity
we do not even need chaotic mixing. Indeed, density perturbations do not grow
in a universe dominated by ultra-relativistic particles if the size of the
universe is smaller than $H^{-1}$. This is exactly what
happens in our model before inflation begins. Therefore the universe should remain relatively
homogeneous until the thermal energy density drops below $V_{dS}$. Once it happens, the universe rapidly becomes exponentially large and homogeneous due to inflation.

The only issue with this scenario is that at the time when the thermal energy density drops below $V_{dS}$, the scalar field should be at the position where the potential $V(\phi)$ is flat and inflation is possible. This is not easy to achieve in the new inflation/hilltop inflation scenario because the probability to land by chance at the tiny vicinity of a maximum of the potential in these models is not very high. However, in the models with long inflationary slopes of the potential, such as the simplest chaotic inflation models \cite{Linde:1983gd} or especially in the models discussed in this paper, where the inflationary slope of the potential is infinitely long, this last part should be very easy.

Thus in this scenario, just as in the simplest chaotic inflation scenario, inflation may begin if we had a sufficiently homogeneous domain of a smallest possible size (Planck scale), with the smallest possible mass (Planck mass), and with the total entropy O(1). We see no reason to expect that the probability of formation of such universes is strongly suppressed.

One can arrive at a similar conclusion from a completely different point of view. Investigation of the quantum creation of a closed or an infinite open inflationary universe with $V\ll 1$ shows that this process is forbidden at the classical level, and therefore only occurs by tunneling. The main reason for this result is that closed de Sitter space always has its size greater than $H^{{-1}} \sim 1/\sqrt V$, and the total energy greater than $H^{-3}V \sim 1/\sqrt V$. Than is why the universe with $V\ll 1$ is large, heavy and difficult to create.  As a result, the probability of this process is exponentially suppressed \cite{Linde:1984ir,Vilenkin:1984wp}. Meanwhile,  creation of  flat or open compact universes is possible without any need for  tunneling, and there would be no exponential suppression for the probability of quantum creation \cite{ZelStar,Coule,Linde:2004nz}. This suggests that the problem of initial conditions for low energy scale inflation can be easily solved if one considers topologically nontrivial compact universes. This  possibility is completely consistent with observations, since inflation lasting more than 60 e-foldings makes all topological effects unobservable.  

Thus we believe that there is no real problem with initial conditions for inflation in a broad class of inflationary models even if inflation may occur only at $V\ll1$. Other ways to solve the problem of initial conditions for low energy scale inflation, including the possibilities related to the string landscape scenario \cite{Eternal,Lerche:1986cx,Bousso:2000xa,Str,Kachru:2003aw,Douglas,Susskind:2003kw}, can be found in \cite{Linde:2014nna}. As we will see now, the cosmological attractor models studied in this paper have additional advantages in this respect.


\section{Cosmological attractors, dS valley, and initial conditions for inflation}\label{ICI}

In order to make our basic idea clear, we would like to start with a discussion of a simple toy model of a massless scalar field with gravity and a cosmological constant $V= const \ll 1$:
 \be
 {1\over \sqrt{-g}} \mathcal{L} = {1\over 2}   R - {1\over 2} {\partial \vp^2}  - V_{dS}   \,  .
\label{CC}\ee
The description of the evolution of the universe in terms of classical space-time begins when the sum of the kinetic and gradient energy of the field $\vp$ and the vacuum energy $V$ for the first time becomes equal to the Planck energy density: ${1\over 2} (\dot\vp^{2}+{\partial_{i} \vp^2})+V_{dS} = 1$. For $V_{dS}\ll 1$, this means that ${1\over 2} (\dot\vp^{2}+{\partial_{i} \vp^2}) \approx 1$.

What happens with such a universe in the future? If it is a compact open or flat universe, then we believe that the answer is already given in the previous section. The field $\phi$ is massless, its perturbations represent relativistic waves/particles, and  because of their ultra-relativistic equation of state combined with the chaotic mixing, the universe stays small and relatively homogeneous, of the size $a(t) \ll H^{{-1}}$. Its size grows as $\sqrt t$ until the energy density of the scalar field $\rho_{\vp}(t) \sim t^{{-2}}$ drops below $V$. After that, the universe becomes an exponentially expanding dS space, which will contain perturbations of the scalar field with an exponentially decreasing amplitude. No other choice is available for the system.

One may consider a more complicated case: the same model \rf{CC}, but the universe was born large from the very beginning. In our opinion,   this assumption  is problematic and unnecessary. If everything that we see now could originate from something small, why should we consider a much more complicated way to reach the same goal? One of the problems addressed by inflation was to explain its huge size.  Since we know how to solve this problem using inflation, an assumption that the  universe was   large prior to inflation is no longer required.  Nevertheless, for the sake of argument, and to cover all options, we will consider this possibility as well. Then there are two basic possibilities: 

1) The whole universe collapses before the onset of exponential expansion in dS space with the cosmological constant $V$.  But this is possible only if the collapse  occurs very quickly, before the energy density of the initially expanding universe becomes smaller than $V_{dS}$ and dS inflation takes over. For $V_{dS} \sim 10^{{-10}}$, this may happen only if the whole universe collapses almost instantly, within about $10^{{-28}}$ seconds from the birth of the universe, which means that we are not talking about a universe with human observers trying to explain its homogeneity, but about a short-living quantum fluctuation. 


2) If we are not talking about a small instantly collapsing ``virtual'' universe, then the next possibility to consider is a large expanding universe with initial inhomogeneities, which may grow and become large in some parts of the  universe, but not everywhere. For example, the universe may contain such local inhomogeneities as primordial black holes. But if the universe as a whole continues to expand, then density of matter in all of its expanding parts rapidly drops down to $V_{dS}$. These parts will continue to grow exponentially, whereas the collapsing parts of the universe will not. As a result, the volume of the universe will become dominated by the exponentially large volume of dS space. Even if it contains black holes or other large inhomogeneous stable structures, they will move away from each other exponentially fast  and therefore eventually they will become irrelevant. This is the similar to the fate that awaits all of us if the dark energy is a cosmological constant: Eventually all distant galaxies and other inhomogeneities will move exponentially far away from us, and the universe will become an empty dS space.

Thus we expect that in the presence of a sufficiently large cosmological constant $V_{dS}$, at least some parts of a large expanding universe do not collapse. The collapsed parts are of little interest to us, whereas all parts that did not collapse within the first $10^{{-28}}$ seconds join the exponential expansion driven by the positive cosmological constant $V_{dS}$. 

In other words, in the context of this simple model, the problem of initial conditions all but disappears. Instead of asking whether it is probable that a large inhomogeneous expanding universe enters a stage of inflation, one may wonder what kind of miracle may prevent at least some  parts of the universe to enter the process of exponential expansion in an eternally growing dS space?

To explain the relation between this toy model of dS space and the theory of initial conditions for inflation, let us take the next step and consider yet another toy model \rf{simpleT2} with an exactly shift-symmetric dS valley potential \rf{PotdS} shown in Fig.~\ref{gorgeexp2}.  This potential represents an infinite dS valley with the shape coinciding with the shape of the dS valley in the simplest T-model \rf{Pot1} in the large $\vp$ limit. According to the argument given above, the universe in the model \rf{simpleT2} should end up at the bottom of dS valley and enter the state of exponential expansion in dS state at $\vt = 0$. 

Note that even after rolling to the dS valley at $\vt = 0$, different parts of the universe may contain different values of the field $\vp$, depending on the initial conditions at the Planck time. 
Indeed, the theory \rf{simpleT2}, \rf{PotdS} is exactly shift-symmetric with respect to the field $\vp$. This means that all initial conditions for the field $\vp$ are equally probable. Therefore all values of this field after it falls to the dS valley are also equally probable: A large dS universe becomes populated with all possible values of the field $\vp$ in its different parts, but
the gradients of this field rapidly become exponentially small due to dS inflation, and the universe becomes locally uniform, up to small inflationary quantum fluctuations of the field $\vp$. 

This is not a complete inflationary scenario yet, because in this model the universe is forever trapped in dS space. In order to implement the standard slow-roll inflation, one should return to realistic models with $Z$-dependent superpotentials discussed in our paper, such as $W = \sqrt{\alpha}  \, \mu \, S \, Z$ proposed  in eq. \rf{simpleT}. These models were  studied in Section~\ref{hyperbolic}. In such models, the field does not stay in dS valley but slowly rolls to the minimum of the potential at small $\vp$. But from the point of view of the theory of initial conditions, these models do not differ much from the toy models of dS space considered above.

Indeed, the kinetic terms in all of these models coincide with the kinetic terms of the theory  \rf{simpleT2}, and the potential of the simplest T-models \rf{Pot1} coincides with the dS valley potential \rf{PotdS} with exponentially good accuracy everywhere except a finite interval of width $|\vp| \lesssim \sqrt\alpha$. According to our results, even if the kinetic energy of the field $\vp$ initially was as large as the Planckian energy, this field does not move by more than $|\Delta\vp| =10$ until it slows down and reaches the slow-roll inflationary regime along the dS valley.   Therefore for an infinitely large range of initial conditions $\sqrt\alpha +O(10)\lesssim |\vp_{0}| < \infty$, the first stages of the cosmological evolution in T-models coincide  with the evolution in the simple dS model \rf{simpleT2}, \rf{PotdS} with exponentially good accuracy. More generally, one can show that  the inflationary regime in this class of models occurs if, after the stage of rolling, the value of the inflaton field at the bottom of the dS valley relaxes at  $|\vp| > \min\bigl[O(\sqrt \alpha), O(1)\bigr]$. This means that the stage of slow-roll inflation occurs for all values of $\vp_{0}$ at the Planck density such that $O(10)\lesssim |\vp_{0}| < \infty$. 

The phase volume of initial conditions with $|\vp| \lesssim  \sqrt\alpha +O(10)$ is finite. Meanwhile the phase volume of all possible inflationary initial conditions in these models is much greater: it is infinitely large, with all values with $|\vp| \gg  \sqrt\alpha +O(10)$ being equally probable because of the shift symmetry. For all values of the field $\vp$ in this infinitely large interval, the probability of the slow-roll inflation in the cosmological attractors studied in this paper coincides with the probability that the universe ends up in an exponentially expanding dS space in the dS valley model \rf{simpleT2}. In accordance with the arguments given in this section, this means that initial conditions for inflation in the family of cosmological attractors studied in this paper are quite natural.

\section{Further generalizations}\label{generalizations}

Before concluding our general investigation of initial conditions for inflation for cosmological attractors, we would like to point out some special cases where this investigation is even simpler, and also to extend our results to other inflationary models.

The first case involves asymmetric T-models \rf{lin}, \rf{expshoulders} shown in Figs. \ref{should} and \ref{gorgeexp}. In these models, for appropriate choice of parameters, the hight of the upper shoulder (i.e. of the dS valley at a higher altitude) can be Planckian. Thus inflation may start at the Planck density, as in the simplest models of chaotic inflation, in which case the problem of initial conditions is easily solved \cite{Linde:1983gd,Linde:1984ir,Vilenkin:1984wp,Linde:1985ub,Linde:1986fd,Linde:2005ht}. When inflation ends at the upper shoulder, the field $\vp$ rolls down to $\vp =0$ and, by inertia, climbs the lower shoulder. As we have already shown, the field cannot move by more than $\Delta \vp \sim 10$. After that, the field stops, and the second (last) stage of inflation begins, which is responsible for the structure formation in the observable part of the universe. We checked that for $\alpha = O(1)$ the duration of the second stage of inflation is more than sufficient to account for the last 60 e-foldings of inflation.

Another interesting possibility emerges in supergravity models with the \K\ potential \rf{NewD} and $\alpha \ll 1/3$, as well as in the models with the \K\ potential \rf{JJ12} and $ \alpha=1/3+\delta$ where $0< \delta  \ll 1/3$. In such models, inflation can begin directly at the Planck boundary with large values of the canonical field $\chi$ in Eq. \rf{expx}. Then the problem of initial conditions for such models is  resolved in exactly the same way as in other versions of chaotic inflation where inflation may start at the Planckian values of the potential $V = O(1)$   \cite{Linde:1983gd,Linde:1984ir,Vilenkin:1984wp,Linde:1985ub,Linde:1986fd,Linde:2005ht}.

The arguments given above can be easily extended to the models discussed in Section \ref{aaa} with the \K\ potential \rf{JJ12}. As we already mentioned, the potential in these models for  $\alpha> 1/3$ also has an infinite dS valley \rf{Pot2} with the $\vp$-independent curvature $m^{2}_{\vt} = 6H^{2}(1-{1\over 3\alpha})$, so most of our arguments apply there for the case $\alpha> 1/3$. The special case $\alpha = 1/3$ is particularly interesting from the point of view of the theory of initial conditions, as already discussed  in \cite{Kallosh:2015zsa}. Indeed, in this case it is convenient to use a different set of coordinates $Z=  e^{i\theta} \, \tanh {\vp\over 2}$ and find that, for the simplest superpotential $W   = \sqrt{\alpha} \,  \mu \ \, S   \,  Z$, the inflaton potential does not depend on the angular variable $\theta$ and is given by $V  = \alpha \mu^2\,  \tanh^{2} {\vp\over \sqrt 2}$. 

\begin{figure}[ht!]
\centering
\includegraphics[scale=.2]{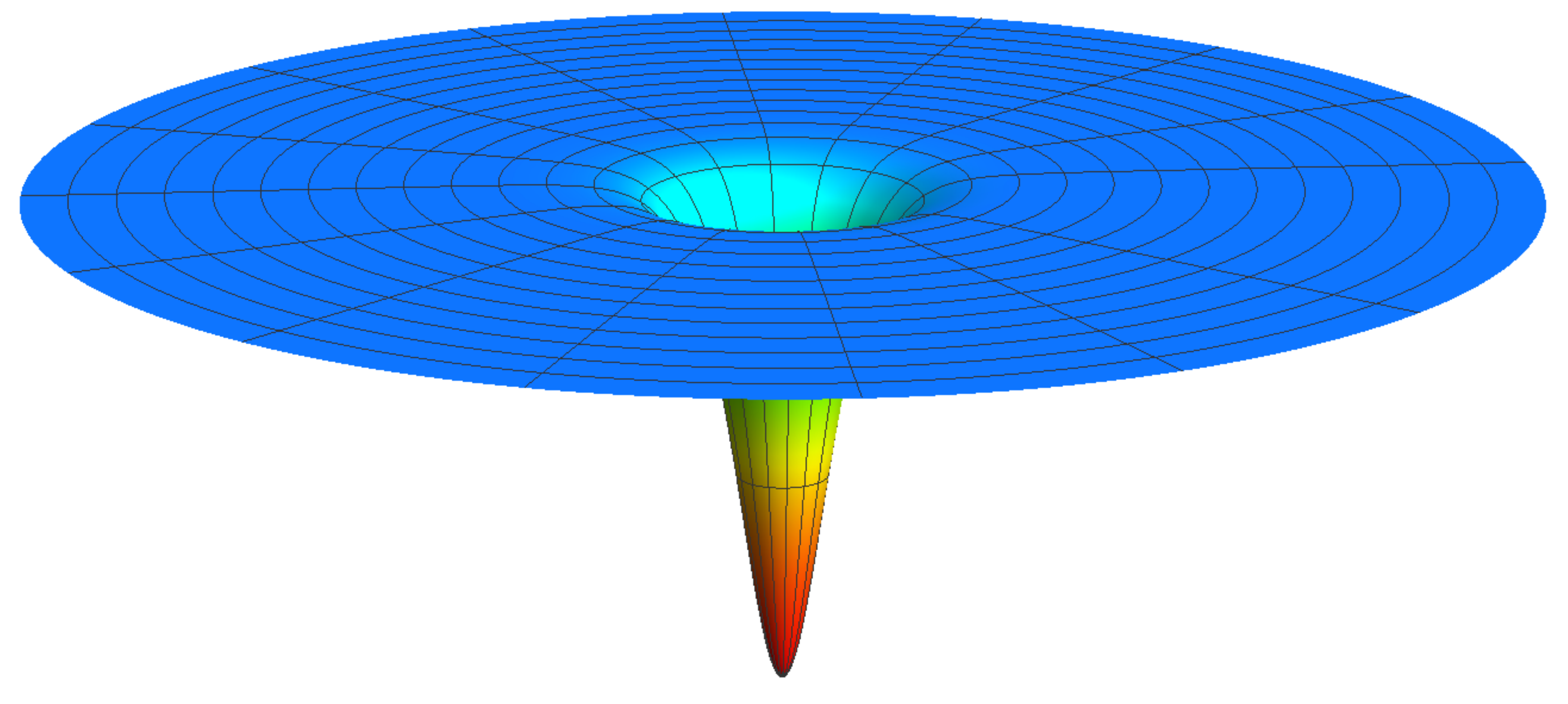}
\caption{\footnotesize The potential $V$ in the theory \rf{JJ12} with $W   = \sqrt{\alpha} \, \mu  \, S   \,  Z$ for $\alpha = 1/3$ in terms of the inflaton field $\vp$ and the angular variable  $\theta$. Everywhere except for the region with  $\vp \lesssim 4$, the potential is flat with exponentially good accuracy, which provides perfect initial conditions for inflation  \cite{Kallosh:2015zsa}. }
\label{gorgeexp3}
\end{figure}

The shape of the potential in these variables is shown in Fig.~\ref{gorgeexp3}. The last 60 e-foldings of inflation in this model are determined by a tiny part of this figure at $\vp \lesssim 4$. The potential $V$ at  large $\vp$ is almost exactly constant, and the phase volume for initial conditions corresponding to $\vp > 4$ is infinitely large. This provides perfect initial conditions for inflation. Independently of the initial velocity of the scalar fields, their kinetic energy rapidly dissipates due to the cosmological evolution. The fields freeze at some point of the infinitely large plateau, until the exponentially slow descent in the radial direction towards the minimum of the potential shown in Fig. \ref{gorgeexp3} begins  \cite{Kallosh:2015zsa}. All of the arguments given above concerning initial conditions for exponential expansion in dS space are immediately applicable to this model.

Up to now, we concentrated on a special class of $\alpha$-attractors in supergravity, describing a combined cosmological evolution of {\it two} fields, $\vp$ and $\vt$. We found that in some exceptional cases the field $\vt$ can help inflation to begin (e.g. in the model \rf{JJ12} with $\alpha \ll 1/3$), but in general it does not play much role in the discussion of initial conditions for inflation: If the field $\vp$ initially was very large, then it falls to the dS valley at very large values of $\vp$ independently of the behavior of the field $\vt$. Our main arguments concerning naturalness of initial conditions for inflation in Section \ref{ICI} take this fact into account. Therefore they equally well apply to all cosmological attractor models based on a {\it single} inflaton field with a plateau dS potential,  such as \rf{cosmo}.  This includes the Starobinsky model \cite{Starobinsky:1980te}, Higgs inflation \cite{Salopek:1988qh},  conformal attractors \cite{Kallosh:2013hoa},  single-field supergravity attractors  \cite{Ferrara:2013rsa}, and universal attractors with non-minimal coupling to gravity \cite{Kallosh:2013tua,Galante:2014ifa}.  More generally, these arguments should apply to any model where the inflaton potential has a sufficiently long and flat slope, as in the simplest models of chaotic inflation \cite{Linde:1983gd}.

\section{Summary}

In this paper we described some simple versions of the theory of the cosmological attractor models \cite{Kallosh:2013hoa,Kallosh:2013yoa,Cecotti:2014ipa,Kallosh:2015zsa, Carrasco:2015uma}. We have used a novel set of coordinates describing the geometry of scalar fields. It  corresponds to a choice of the Killing adapted geometry where the inflaton shift symmetry,  $\vp \rightarrow \vp \, +$ const  of the \K\, potential,  is manifest and is only broken by the superpotential. As the result, the potential of these refined models has an interesting universal property: at large values of the inflaton field $\vp$ it looks like an infinite dS valley of nearly exactly constant depth and width, see Fig. \ref{tr}. In other words, at large $\vp$ the potential has shift symmetry with respect to the shift of the field $\vp$. This property of the theory is not immediately apparent from its formulation, but can be established using the properties of the hyperbolic geometry of the moduli space discussed in \cite{Kallosh:2015zsa,Carrasco:2015uma} and further studied in our paper. The novel element contained in this paper is the choice of conformally flat coordinates, and also a particular choice of the \K\ potential,   which makes the shift symmetry manifest and ensures stability of the dS extremum.

We believe that this interesting property is sufficient for solving the problem of initial conditions in this class of inflationary models. This can be done in several different ways discussed in \cite{Linde:2014nna} and  in Section \ref{review} of our paper, but the existence of the shift symmetry revealed in this paper allows for yet another solution which is rather simple and general.

Our investigation shows that during the pre-inflationary stage of expansion of the universe, the canonically normalized inflaton field $\vp$ cannot change by more than $O(10)$. Therefore for all initial conditions for the field $\vp$, except for a finite interval near the minimum of the potential at $\vp = 0$, the fields roll down to the lower part of the dS valley far from $\vp = 0$, and inflation almost inevitably begins there. This argument is directly applicable as well to all cosmological attractor models, such as \rf{cosmo}, based on a single real inflaton field $\phi$ with a plateau potential, and to any other model where the inflaton potential has a sufficiently long and flat slope. 

\subsection*{\bf Acknowledgements} The authors are grateful to D. Roest for the ongoing collaboration on the cosmological attractors, to S. Ferrara for enlightening comments, and to T. Abel, W. East, M. Kleban,  L. Senatore and J. Zrake for stimulating discussions of the problem of initial conditions for inflation. JJMC, RK and AL are supported by the SITP and by the NSF Grant PHY-1316699. JJMC and RK are also supported by the Templeton foundation grant `Quantum Gravity Frontiers,' and AL is supported by the Templeton foundation grant `Inflation, the Multiverse, and Holography.'   While JJMC attended the KITP program on `Quantum Gravity Foundations: UV to IR' his research was supported in part by the National Science Foundation under Grant No. NSF PHY11-25915.

\end{document}